\documentclass[%
aip,
jcp,%
amsmath,amssymb,
reprint,%
]{revtex4-1}
\usepackage{color}

\newcommand{\sub}[1]{_{\mbox{\scriptsize {#1}}}}

\usepackage{graphicx}
\usepackage{dcolumn}
\usepackage{bm}

\def\siml{\hspace{1ex} ^{<} \hspace{-2.5mm}_{\sim} \hspace{1ex}}

\begin{document}

\preprint{AIP/123-QED}

\title[Large Scale MD simulations of Nucleation]{Large Scale Molecular Dynamics Simulations of Homogeneous Nucleation}

\author{J\"urg Diemand}
\email{diemand@physik.uzh.ch}
\homepage{http://www.physik.uzh.ch/~diemand/}
\author{Raymond Ang\'elil}%
\affiliation{Institute for Theoretical Physics, University of Zurich, 8057 Z\"urich, Switzerland}

\author{Kyoko K. Tanaka}
\author{Hidekazu Tanaka}
\affiliation{Institute of Low Temperature Science, Hokkaido University, Sapporo 060-0819, Japan}

\date{\today}

\begin{abstract}
We present results from large-scale molecular dynamics (MD) simulations of homogeneous vapor-to-liquid nucleation.
The simulations contain between one and eight billion Lennard-Jones (LJ) atoms, covering up to 1.2 $\mu$s (56 million time-steps).
They cover a wide range of supersaturation ratios, $S \simeq 1.55$ to $10^4$, and temperatures from $kT = 0.3$ to $1.0\epsilon$
(where $\epsilon$ is the depth of the LJ potential, and $k$ the Boltzmann constant).
We have resolved nucleation rates as low as $10^{17}$ cm$^{-3}$ s$^{-1}$ (in the argon system),
 and critical cluster sizes as large as 100 atoms. Recent argon nucleation experiments probe nucleation rates in an overlapping range,
making the first direct comparison between laboratory experiments and molecular dynamics simulations possible: 
We find very good agreement within the uncertainties, which are mainly due to the extrapolations of argon and LJ saturation curves to very low temperatures.
The self-consistent, modified classical nucleation model of Girshick and Chiu [J. Chem. Phys. 93, 1273 (1990)]
underestimates the nucleation rates by up to 9 orders of magnitudes at low temperatures,
and at $kT = 1.0\epsilon$ it overestimates them by up to $10^5$. 
The predictions from a semi-phenomenological model by Laaksonen et al. [Phys. Rev. E {\bf49}, 5517 (1994)]
are much closer to our MD results, but still differ by factors of up to $10^{4}$ in some cases.
At low temperatures, the classical theory predicts critical clusters sizes, which match the simulation results (using the first nucleation theorem) quite well, while the
semi-phenomenological model slightly underestimates them. At $kT = 1.0\epsilon$ the critical sizes from both models are clearly too small.
In our simulations the growth rates per encounter, which are often taken to be unity in nucleation models, lie in a range from $0.05$ to $0.24$.
We devise a new, empirical nucleation model based on free energy functions derived from subcritical cluster abundances,
and find that it performs well in estimating nucleation rates.
\end{abstract}

\pacs{05.10.-a, 05.70.Fh, 05.70.Ln, 05.70.Np, 36.40.Ei, 64.60.qe, 64.70.Hz, 64.60.Kw, 64.10.+h, 83.10.Mj, 83.10.Rs, 83.10.Tv}
\keywords{drops, droplets, Lennard-Jones potential, molecular dynamics method, nano-clusters, nucleation, phase transitions, solid-vapor transformations}
\maketitle

\section{Introduction}

The first order phase transition from vapor to liquid via homogeneous nucleation is a ubiquitous fundamental process and plays an important roles in many areas of science and technology. Despite the 
familiarity of the process, serious unreliability remains in model predictions for nucleation rates, because the surface properties of the small droplets are poorly understood\citep{Kalikmanov2013}. 

The widely used classical nucleation theory (CNT) \cite{volmer, becker, zeldovich, feder} estimates the work required to form liquid droplets  under the assumption that they 
resemble bulk liquid, have a sharp boundary, as well as the same surface tension as macroscopic drops. However, the smallest stable, or critical, clusters are nano-sized at typical nucleation conditions, and their
properties differ significantly from the CNT assumptions. This results in massive discrepancies between the nucleation rates predicted by CNT and those measured in laboratory experiments \cite{schmitt1, schmitt2, adams, wright, viisanen,viisanen2, anisimov,Tanaka2011, kalikmanovReview, napari}
and molecular dynamics\cite{hale, laasonenWoo, frenkel, wolde, chkonia, Tanaka2005,Wedekind2007,Tanaka2011, yasuoka,yasuoka2, toxvaerd2, toxvaerd1} or Monte Carlo simulations\cite{kusaka, oh, senger, wolde:1591, oh2, chen,yoo, merikantoMC}.

More recent nucleation theories have made significant improvements since the introduction of the CNT.
Density functional theory (DFT)\cite{DFT1} and the extended modified liquid drop model\cite{} take into account the extended transition region from liquid to vapor, sometimes referred to as the the ``corona",
and match MD results far better than the CNT\cite{}. Here we use MD simulations to test
another approach: the semi-phenomenological (SP) model\cite{SPmodel,delale,ford1993,Laaksonen1994,KalikmanovSPmodel}, which corrects the cluster formation energy from CNT
by using the second virial coefficient. The SP model agrees well with experimental data on water, nonane and n-alcohols,\cite{delale} and also with MD simulations of Lennard-Jones atoms
at high supersaturations \cite{Tanaka2005,Tanaka2011}. However, the range of applicability of the SP model remains unclear: the scaling law proposed by McGraw and Laakonsen\cite{mcgraw}
presents a different correction to the CNT, which is supported by DFT calculations\cite{mcgraw} and Monte Carlo simulations\cite{merikantoMC} for clusters greater than a certain size. This suggests that the SP model
becomes inapplicable at low supersaturations, where the critical clusters are larger.
 
\begin{figure*}
\includegraphics[height=.55\textheight]{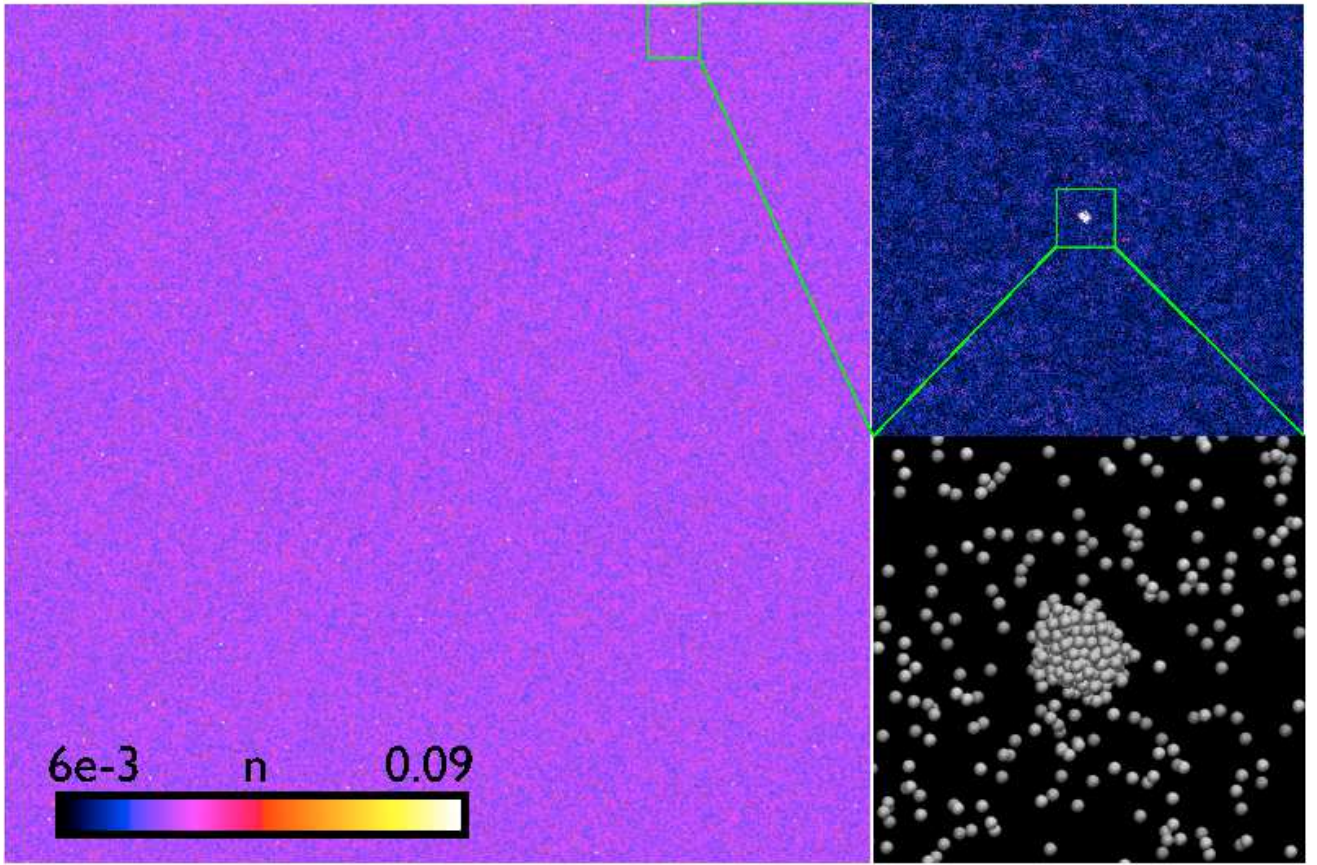}
\caption{A snapshot taken at the end of run T6n8. The left panel shows a slice with a depth of  $300 \sigma$ of the entire box, $5000 \sigma \times 5000 \sigma$.
The insert in the upper-right is $300 \sigma \times 300 \sigma \times 300 \sigma$ . The colour map represents density. The final insert in the bottom right is $40 \sigma \times 40 \sigma \times 20 \sigma$ . The cluster in view here has 220 members.}\label{snapshot}
\end{figure*}

Molecular dynamics simulations are able to directly resolve details of the nucleation process, and provide useful test cases for nucleation models \cite{Tanaka2005,Tanaka2011}.
The size of the simulations - the number of atoms and time-steps - determines the nucleation rates that can be resolved. Typical MD simulations of homogeneous nucleation use $10^4$ to $10^5$ atoms. An exception to this are recent simulations with up to $10^6$ atoms\cite{Horsch&Vrabec2009}.
At low vapor densities, and therefore low supersaturations, a single nucleation event becomes unlikely to occur within reasonable computational timeframes. Large, distributed simulations however, allow for the occurrence of such rare nucleation events, and enable us to measure these low rates. 
Here we present results from very large scale MD simulations with
between one and eight billion Lennard-Jones atoms, evolved over ranges from 250 thousand up to 56 million time-steps. Figure \ref{snapshot} is a snapshot taken towards the end of one of our simulations.

Most laboratory nucleation experiments are carried out at relatively low
supersaturations, and measure nucleation rates $J$ less than
$10^{10}$cm$^{-3}$s$^{-1}\;$[\cite{schmitt1, schmitt2, adams, wright, viisanen,viisanen2, anisimov}].  Recent development in
Supersonic Nozzle (SSN) nucleation experiments has increased the
accessible rates enormously\cite{sinha}.  For the case of Argon, SSN experiments
in the temperature range from 34 to 53 K resolve 
nucleation rates of $10^{17}$cm$^{-3}$s$^{-1}$.  In comparison,
current MD simulations probe $J$ values in a regime well above
$10^{21}$cm$^{-3}$s$^{-1}$ [\cite{Tanaka2005,Wedekind2007,Tanaka2011}].
Our large scale simulations manage to bridge this gap,  making a direct comparison between
simulations and experiment possible.

Our simulations are a direct extension to lower supersaturations of the
recent studies by Tanaka et al. \cite{Tanaka2005,Tanaka2011}. The larger particle numbers offer several advantages: 
\begin{enumerate}
\item Resolving and quantifying nucleation at low supersaturations becomes possible within an accessible number of simulation time-steps, despite the rather slow nucleation process. 
\item Even after forming many stable droplets, the vapor depletion is negligible: The supersaturation remains effectively constant throughout the simulations.
\item Excellent statistics on liquid droplet abundances and their microscopic properties can be obtained, such as density profile, shape, and surface and core atom potential energies over a wide range of droplet sizes. These results are to be presented in a subsequent paper.  (Ang\'elil et al. in preparation)
\item Since the number of particles is very large and the computational volume is much larger than the force cutoff, large scale simulations can be run very efficiently on a large number of processor cores. 
\item Because so few clusters are formed relative to the number of atoms in the gas, the amount of temperature rescaling necessary to maintain the average temperature at a constant level is minimal. We therefore need not worry about artificial thermostatting effects biasing the simulation results. 
\end{enumerate}

Section \ref{sec:theory} provides a concise summary of the CNT, the modified CNT (MCNT) and the SP model. 
Section \ref{sec:simulations} describes our MD simulations, in section \ref{sec:results} we present our results and in section \ref{sec:hybrid} we introduce a new empirical nucleation model based on the subcritical equilibrium cluster abundances from the simulations. Finally, section \ref{sec:summary} concludes the paper by summarising our findings. 

\section{Theoretical models for homogeneous nucleation}\label{sec:theory}

In this section the theoretical models used in this work are summarised briefly, for more details see e.g.\cite{Tanaka2011,Kalikmanov2013}.
The free energy $\Delta G(i)$ associated with forming a liquid cluster of size $i$ from the vapor phase has a positive surface term, corresponding
to the work required to form the vapor-liquid interface, and a volume term which is negative for supersaturated vapor. $\Delta G(i)$ reaches a maximum at
a critical cluster size $i^*$. Larger clusters are considered to be stable and smaller ones unstable. The equilibrium number density of small, unstable clusters is
\begin{equation}\label{eqdist}
n_e(i) = \frac{P_1}{kT} \exp \left[ - \frac{\Delta G(i)}{kT} \right] \;, 
\end{equation}
where $P_1$ is the monomer pressure. For simplicity clusters are assumed to grow and shrink by accretion and evaporation of monomers only,
which often is an accurate assumption because monomers are usually the most abundant species.
The total growth rate is now
\begin{equation}\label{growth}
\frac{di}{dt} = R^+(i) - R^-(i)  \; ,
\end{equation}
where the accretion rate $R^+(i)$ is the transition rate from i-mer to (i+1)-mer per unit time. The evaporation rate
$R^-(i)$ is the transition rate from i-mer to (i-1)-mer per unit time. $R^+(i)$ is given by
\begin{equation}\label{rplus}
R^+(i) = \beta n_e(1) \nu_{\rm th} \; 4\pi r_0^2 i^{2/3}  \;,
\end{equation}
where the sticking probability $\beta$ is the probability that a monomer which encounters a cluster of size $i$ is accreted.
$\nu_{\rm th}$ is the mean thermal velocity. $r_0$ is the mean inter-particle separation in the liquid phase,
so $4\pi r_0^2 i^{2/3}$ corresponds to the surface area of an $i$-mer. 
We use the total growth rates measured in the simulations to define $\alpha$, a growth rate per encounter: 
\begin{equation}\label{alpha}
\alpha \equiv \frac{di/dt}{n_e(1) \nu_{\rm th} \; 4\pi r_0^2 i^{2/3} } = \frac{3}{n_e(1) \nu_{\rm th}  4\pi r_0^2}\frac{d \left(i^{1/3}\right)}{dt} \; ,
\end{equation}
so that $\alpha = 1$ means growth at the kinetic rate. 
For large clusters and supersaturations one can neglect evaporation ($R^-(i) \simeq 0$), the growth rate is the same as  the accretion rate and the
growth rate per encounter $\alpha$ equals the sticking probability $\beta$.
Including a (model dependent) evaporation term $R^-(i)$ one finds for large clusters (see Appendix \ref{sec:evaporation}):
\begin{equation}\label{beta}
\frac{di}{dt}  \simeq R^+(i) \left[ 1 - \frac{1}{S}\right]  = \beta n_e(1) \nu_{\rm th} \; 4\pi r_0^2 i^{2/3} \left[ 1 - \frac{1}{S}\right] 
\end{equation}
where $S \equiv P / P_{\rm sat}$ is the supersaturation ratio. Eq. (\ref{beta}) suggests that evaporation becomes significant at low supersaturations,
even for large, stable clusters. Classical nucleation models usually assume $\alpha=1$. In Section \ref{sec:sticking}
we use Eqs. (\ref{alpha}) and (\ref{beta}) to determine the actual $\alpha$ and $\beta$ from the growth rates observed in our MD simulations.

The nucleation rate $J$ is approximately proportional to the abundance of critical clusters and their transition rate \cite{Oxtoby,Tanaka2011,Kalikmanov2013}:
\begin{equation}\label{eq:j}
J =  \left[  \sum_{\rm i=1}^{\infty} \frac{1}{R^+(i) n_e(i) }\right]^{-1} \simeq R^+(i^*) n_e(i^*) Z \;, 
\end{equation}
where Z is the Zeldovich factor
\begin{equation}
Z  = \sqrt{ \frac{-1}{2\pi k T} \frac{d^2 \Delta G(i^*)}{di^2}  }\;.
\end{equation}

In the classical nucleation theory (CNT)\cite{Oxtoby} the free energies $\Delta G(i)$ are assumed to follow
\begin{equation}
\frac{\Delta G_{\rm CNT}}{kT} = -i \ln S + \eta i^{2/3} \label{CNT} \; , 
\end{equation}
where the surface term has a pre-factor of
\begin{equation}\label{eq:eta}
\eta = \frac{4\pi r_0^2\gamma}{kT} 
\end{equation}
and $\gamma$ is the condensed phase's planar surface tension. $\Delta G_{\rm CNT}$ peaks at the critical cluster size
\begin{equation}\label{icrit}
i^*_{\rm CNT} = \left[  \frac{2}{3} \frac{\eta}{\ln S} \right]^{3} \; ,
\end{equation}
and the classical nucleation barrier is
\begin{equation}\label{barrier}
\frac{\Delta G_{\rm CNT} (i^*)}{kT} =  \left[  \frac{4}{27} \frac{\eta^3}{(\ln S)^2} \right]  \; .
\end{equation}

$\Delta G(i=1)$ must be zero to get the correct $n_e(1)$ from Eq. (\ref{eqdist}).
Therefore several authors \cite{girshick1990kinetic,Oxtoby,Ford1997} subtract a constant from $\Delta G$ to arrive at a modified (or self-consistent) CNT, referred to as
MCNT hereafter:
\begin{equation}
\frac{\Delta G_{\rm MCNT}}{kT} = -(i-1) \ln S + \eta  (i^{2/3}-1) \label{MCNT} \;. 
\end{equation}
The critical sizes of CNT and MCNT are the same and given by Eq. (\ref{icrit}).

A semi-phenomenological (SP) model was proposed by Meier and Dillman\cite{SPmodel} and then developed further in several studies\cite{delale,ford1993,Laaksonen1994,KalikmanovSPmodel}.
Here we use the version presented in Laaksonen et al. \cite{Laaksonen1994}. The SP model adds one extra term to $\Delta G$:
\begin{equation}
\frac{\Delta G_{\rm SP}}{kT} = -(i-1) \ln S + \eta  (i^{2/3}-1)  +  \xi (i^{1/3} -1 )\label{SP} 
\end{equation}
and the extra parameter $\xi$ is fixed so that the formation energy of a dimer $\Delta G(i=2)$ agrees with
the value derived from the second virial coefficient $B_2$. See Appendix \ref{sec:model_details} and Table (\ref{tab:t2}) for details.

Note that we define the supersaturation $S \equiv P / P_{\rm sat}$ using the total pressures throughout this paper.
In the theoretical models, the supersaturation actually refers to the ratio of monomer pressures $P_1 / P_{\rm sat,1}$. 
At low temperatures the resulting supersaturations are nearly identical. At $kT = 1.0 \epsilon$ the monomer saturation ratios are quite different, and they even
fail to rise with increasing total number density, total pressure and nucleation rate:
the highest $P_1 / P_{\rm sat,1}$ is actually found for run T10n60 with 1.53. It is higher than in our highest number density, highest nucleation rate simulation T10n62,
where the lower monomer abundance leads to a $P_1 / P_{\rm sat,1}$ of only 1.50. To avoid this problem we use the total
pressures to define the supersaturation $S$ everywhere.

Section \ref{sec:results} compares the nucleation rates predicted by MCNT and the SP model to those obtained from MD.
Section \ref{sec:hybrid} uses elements from theory as well as data from MD in an attempt to estimate the free energy; and from this, the nucleation rate.

\section{Numerical Simulations}\label{sec:simulations}

\subsection{Simulation code, setup and parameters}

The simulations were performed with the Large-scale Atomic/Molecular Massively Parallel Simulator (LAMMPS) code\cite{LAMMPS}, developed at Sandia National Laboratories and distributed as
open source code. It is a highly optimised, widely used and well tested code. We have confirmed that it reproduces results from earlier
nucleation simulations obtained with independent codes\cite{Tanaka2005,Wedekind2007,Tanaka2011}.
Using message passing and spatial domain decomposition, LAMMPS is able to run efficiently on very large supercomputers.
Due to the large number of atoms in a relatively homogeneous configuration, and due to the short range of the interactions, the simulations described here scale extremely well with
processor core count.  We are able to run one billion particle simulations on 32'768 cores on the HERMIT and SuperMUC supercomputers,
at 88 to 95 \% efficiency relative to running with the same number of atoms on only 1024 cores.

We use the Lennard-Jones potential 
\begin{equation}\label{lj}
\frac{u(r)}{4\epsilon} = \left(\frac{\sigma}{r}\right)^{12} - \left(\frac{\sigma}{r}\right)^6,
\end{equation} 
except cut-off and shifted to zero at $5\sigma$. The thermodynamic properties of  the LJ fluid depend on the cutoff scale \cite{psattot,Napari2001,Dunikov2001,baidakov}. The scale of $5\sigma$ is widely used in nucleation simulations
\cite{Tanaka2005,Wedekind2007,Kraska2006,Tanaka2011} and the resulting fluid comes relatively close to the full potential LJ-fluid and to real argon \cite{Mecke1997, psattot,Napari2001,Dunikov2001,baidakov}
at a reasonable computational cost. In Section \ref{sec:conv} we explore the effects of increasing the cutoff scale to $6.78\sigma$.

The simulation box has periodic boundary conditions. As clusters form, the total potential energy drops and so in a constant energy system, the temperature would increases.
We force the average temperature to be constant by simply rescaling
the velocities at every time-step. In nucleation simulations, this simple method gives the same results as the use of a carrier gas for temperature control or other more sophisticated thermostat algorithms\cite{Tanaka2005, wedekind2007influence}.
In the large-volume-low-nucleation-rate simulations presented here, the required amount of rescaling turns out to be extremely small: We can even turn off the velocity rescaling and still find very
similar nucleation rates, see Section \ref{sec:conv}.

We use the standard velocity-Verlet (also known as leap-frog) integrator and the time-steps are set to $\Delta t = 0.01 \tau = 0.01 \sigma\sqrt{m/\epsilon},$ considerably less than the oscillation time $\tau$. The soundness of this time step has been verified through convergence tests, see Section \ref{sec:conv} and also here\cite{Tanaka2005}. In the argon system the units are $\epsilon/k = 119.8 $K, $\sigma=3.405$\AA, $m= 6.634 \times 10^{-23}$g and $\tau = 2.16 $ps.

\subsection{Initial conditions}

The initial conditions are random positions and velocities from a pseudorandom number generator with a sufficiently large period, and high statistical quality \cite{random_generator}. The random positions contain some highly overlapping atoms which lead to unrealistically strong repulsive Lennard-Jones forces during the first few time-steps.
To limit the effects of such artificially high accelerations, the particle velocities are limited to 0.1 $\sigma/ \Delta t$ = 10.0 $\sigma/ \tau$ (or $\sim1600$ $m/s$ for Argon), which is at least  6.4 times
higher than the mean thermal velocities in all our simulations. Starting from simple cubic grid initial positions instead, gives the same results (Section \ref{sec:conv}).
The properties used to set up the simulations are given in table \ref{tab:t1}.

\subsection{Analysis}

\begin{figure}

\includegraphics[height=.3\textheight]{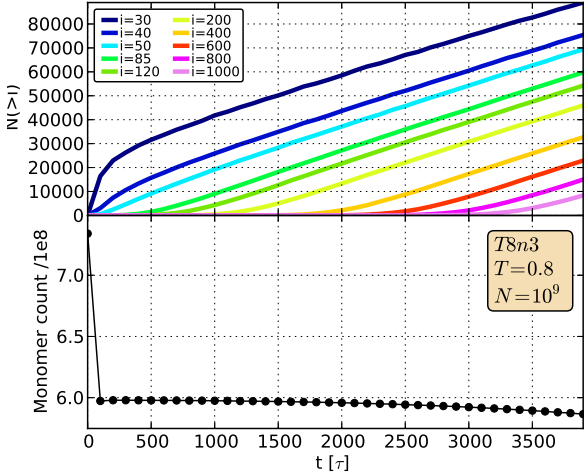}
\caption{Top: Number of clusters above various threshold sizes as a function of time for run T8n3. Bottom: Evolution of monomer count
for the same run.
This is our highest nucleation rate run, and the only run in which the monomer depletion fraction is significant by the end of the run.}\label{fig:T8n3}
\end{figure}

Liquid clusters are defined using the Stillinger criterion\cite{Stilinger}, which iteratively joins atoms with small enough separations $r < r_c$ 
into a common group. We adopt the same temperature-dependant linking lengths $r_c$ as previous studies\cite{Tanaka2005,Tanaka2011}. These are listed in Table \ref{tab:t2}. The choice of cluster definition has some effect on some of our results (for example to size distributions)\cite{wedekind2007best}, while the nucleation rates are not affected, because they do not depend on the absolute cluster sizes.

Liquid clusters are identified on-the-fly and recorded many times during each run. The results described here are all based on these cluster counts. The simulations also
provide more detailed, microscopic information about the liquid clusters; such as their accretion and evaporation rates, density profiles, shapes and binding energies; along with excellent
statistics over a wide range of cluster sizes. These properties will be presented in an upcoming publication (Ang\'elil et al., in preparation).

Figure \ref{fig:T8n3} illustrates the evolution of cluster counts and number of monomers for a relatively high temperature, high nucleation rate case.
The number of monomers, as identified by the group-finding algorithm, in the initial conditions is
smaller than $N=10^9$ because of the overlap in the random initial positions. Over the first 100$\tau$ of the simulation, the monomer count falls off rapidly as the equilibrium distribution
(see Eq. \ref{eqdist}) of small, unstable clusters takes shape. The SP and MCNT models predict the critical cluster size $i^*$ to be between 20 and 25. Subsequently, we observe a perfectly linear increase in $N(>i)$, which is the number of clusters larger than some threshold size $i$. Note how small the decreases in monomer count (and therefore pressure and supersaturation) are during this nucleation phase, even though large amounts of stable clusters are
produced. This allows us to follow the nucleation process and cluster growth in a realistic, nearly-constant-pressure environment; without significant depletion of the vapor.
Our typical runs undergo significantly less monomer depletion than this relatively high $T$, high $J$ case. 

We use the Yasuoka-Matsumoto method\cite{yasuoka} (also referred to as threshold method) to measure the nucleation rate:
$J$ is given by the slope of a least squares linear fit to $N(>i)$. As evident in Figure \ref{fig:T8n3}, the good statistics of this run
allow us to measure the nucleation rate especially precisely. Furthermore, we arrive at the same nucleation rate over a wide range of threshold sizes, as seen in earlier nucleation simulations\cite{matsubara}.  For the linear fits, the initial lag time must be ignored. This simply reflects the time needed for the quasi-steady state gas to fully form (finally resulting in the distribution of subcritical clusters), and also for stable clusters to grow to a certain size. 
More sophisticated analysis methods would allow to fit also the lag time and the transition period\cite{Shneidman1999}, but here we focus our analysis on the much simpler steady-state regime.

\subsection{Numerical convergence tests}\label{sec:conv}

\begin{figure}\label{fig:conv}
\includegraphics[height=.4\textheight]{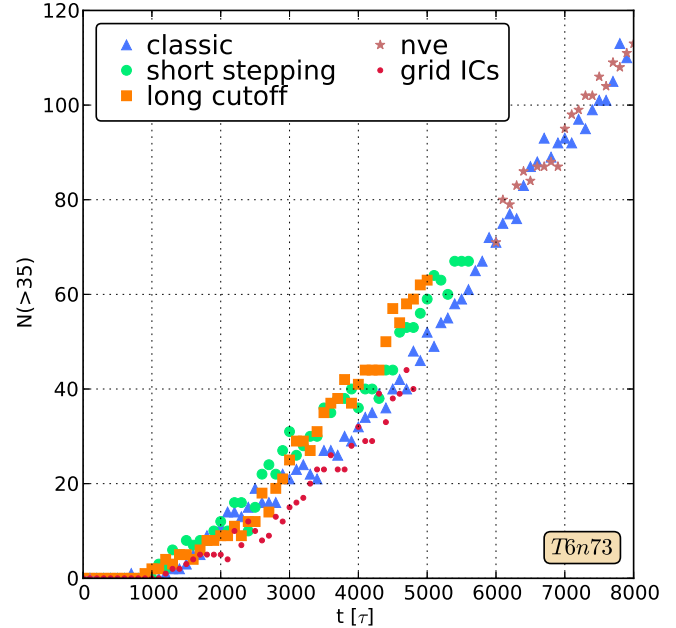}
\caption{Number of clusters with more than 35 member as a function of time for run T6n73 (blue triangles) and 4 more runs at the same number density and temperature, except with different algorithmic choices.}
\end{figure}

To assess the impact of our chosen numerical parameters on the measured nucleation rates we performed four additional simulations with the same physical properties as run T6n73. In each one of these
additional simulations one of the numerical parameters was varied significantly from the standard setup described above. They explore the effect of shorter time-steps, a longer force cutoff, starting from a regular grid instead of random initial conditions and turning off the velocity rescaling (i.e. NVE instead of NVT). We find that only the longer force cutoff changed the measured nucleation rates measurably: Going from our standard $5\sigma$ cutoff to a $6.78\sigma$ cutoff increases the nucleation rates by about 13\%, see Figure \ref{fig:conv}.

Run T6n73NVE was started at time $\tau = 6'000$ using the restart file from run T6n73. It was run until $\tau = 8'000$, i.e. for a period of $\tau = 2'000$, which corresponds to 200'000 time-steps. No velocity rescaling was performed in run T6n73NVE, it represents a micro-canonical or NVE ensemble (number of particles N, volume V and energy E are constant). NVE simulations of nucleation have been presented in Kraska (2006)\cite{Kraska2006}, the high nucleation rates probed ($J>10^{25}$ cm$^{-3}$s$^{-1}$) lead to strongly increasing average temperatures due to the latent heat from condensation. Our run T6n73NVE has a much lower nucleation rate and the resulting temperature increase is tiny: over the entire run period ($\tau = 2'000$), the average temperature did increase from $T=0.6$ to $T=0.600026$, a relative increase of $5.0\times 10^{-5}$. The measured nucleation rate agrees with our fiducial NVT simulation within the uncertainty of  a few percent in the slope of the linear fit. This implies, that after some initial equilibration period, the velocity rescaling has only very minor effects in our simulations and we would obtain very similar results without this somewhat artificial and unphysical velocity rescaling. 

The total energy is conserved very accurately in run T6n73NVE: During the entire run the relative energy deviation form the initial value, $|E(t) - E_0|/E_0$, remains less than $5.0 \times 10^{-8}$. The accurate energy conservation and previous convergence tests\cite{Tanaka2005} indicate, that our fiducial time-step of $\Delta t = 0.01 \tau$ is sufficient. For an explicit test, we run T6n73 with five times shorter time-steps of $\Delta t = 0.002 \tau$. The evolution of the number of stable clusters is similar as in the fiducial run T6n73 (see Figure \ref{fig:conv}) and the resulting nucleation rate agrees perfectly within the uncertainties of the slope estimates.

Throughout this work we use an LJ-potential with a cutoff distance of $5\sigma$, which is widely used in nucleation simulations\cite{Tanaka2005,Wedekind2007,Kraska2006,Tanaka2011}.
The thermodynamic properties of  the LJ fluid depend on the cutoff scale, especially the surface tension of a fluid with a $5\sigma$ cutoff lie
a few percent below the $6.78\sigma$ cutoff values and the full potential values \cite{psattot,Napari2001,Dunikov2001,baidakov}. For the comparisons with theoretical models, we use surface tension values from \cite{baidakov}, who used a $6.78\sigma$ cutoff. To check the influence of cutoff scales on nucleation rates we run a simulation identical to run T6n73, but with a cutoff scale of $6.78\sigma$ instead of $5\sigma$. The resulting nucleation rates are similar, the longer cutoff gives
a 13 percent higher nucleation rate. This small difference does not affect our model comparisons, where we discuss much larger differences, often several orders of magnitude. The slightly higher rate in the simulation with a larger cutoff is surprising, since 
surface tension is expected to increase with the cutoff scale\cite{Napari2001}, which should lead to a larger classical nucleation barrier (Eq. \ref{barrier}) and therefore a lower nucleation rate. Detailed numerical confirmation and further study of
the influence of cutoff scale on nucleation rates would be worthwhile, but are beyond the scope of this work.
 
\begin{table}
\caption{Simulation properties: temperature $T$, number of atoms $N$, periodic cube size $L$, atom number density $n$ and total run time.}
\begin{ruledtabular}
\begin{tabular}{ l c c c c c }
Run ID&T&N&L& $n_{t=0}$& $t_{\rm{end}}$    \\
& [$\epsilon/k$] & & [$\sigma$] & $\left[\sigma^{-3}\right]$ & $\left[\tau \right]$
\\
\hline
  T10n62 & 1.0 & $10^9$ & 2526.6 & $6.20\times10^{-2}$ & $1.28 \times 10^{3}$ \\

  T10n60 & 1.0 & $10^9$ & 2554.4 & $6.00\times10^{-2}$ & $2.55 \times 10^{3}$ \\

  T10n58 & 1.0 & $10^9$ & 2583.4 & $5.80\times10^{-2}$ & $9.33 \times 10^{3}$ \\ 
   
  T10n55 & 1.0 & $10^9$ & 2629.5 & $5.50\times10^{-2}$ & $2.37 \times 10^{4}$ \\ \hline
  
  T8n30 & 0.8 & $10^9$ & 3218.3 & $3.00\times10^{-2}$ & $3.98 \times 10^{3}$ \\
  
  T8n25 & 0.8 & $10^9$ & 3420.0 & $2.50\times10^{-2}$ & $4.03 \times 10^{3}$ \\
   
  T8n23 & 0.8 & $10^9$ & 3420.0 & $2.30\times10^{-2}$ & $5.60 \times 10^{3}$ \\
   
  T8n20 & 0.8 & $10^9$ & 3684.0 & $2.00\times10^{-2}$ & $1.13 \times 10^{5}$ \\\hline

  T6n80 & 0.6 & $10^9$ & 5000.0 & $8.00\times10^{-3}$ & $5.00 \times 10^{3}$ \\

  T6n73 & 0.6 & $10^9$ & 5155.0 & $7.30\times10^{-3}$ & $8.00 \times 10^{3}$ \\

  T6n65 & 0.6 & $10^9$ & 5358.3 & $6.50\times10^{-3}$ & $3.00 \times 10^{4}$ \\
  
  T6n55 & 0.6 & $10^9$ & 5848.0 & $5.00\times10^{-3}$ & $1.81 \times 10^{5}$ \\\hline

  T5n40 & 0.5 & $10^9$ & 5000.0 & $4.00\times10^{-3}$ & $4.20 \times 10^{3}$ \\

  T5n32 & 0.5 & $10^9$ & 5358.3 & $3.20\times10^{-3}$ & $9.00 \times 10^{4}$ \\
  
  T5n26 & 0.5 & $10^9$ & 5848.0 & $2.60\times10^{-3}$ & $2.45 \times 10^{5}$ \\\hline

  T4n10 & 0.4 & $10^9$ & 10000 & $1.00\times10^{-3}$ & $3.95 \times 10^{4}$ \\
 
  T4n7 & 0.4 & $10^9$ & 11263 & $0.70\times10^{-3}$ & $2.85 \times 10^{5}$ \\

  T4n6 & 0.4 &8$\times$10$^9$& 23713& $0.60\times10^{-3}$ & $2.70 \times 10^{4}$ \\

  T4n5 & 0.4 & $10^9$ & 12599& $0.50\times10^{-3}$ & $5.61 \times 10^{5}$ \\\hline

  T3n14 & 0.3 & $10^9$ & 19259 & $1.40\times10^{-4}$ & $1.55 \times 10^{5}$ \\

  T3n12 & 0.3 & $10^9$ & 20274 & $1.20\times10^{-4}$ & $1.90 \times 10^{5}$ \\
  
  T3n9 & 0.3 & $10^9$ & 22314 & $0.90\times10^{-4}$ & $3.75 \times 10^{5}$ \\
\end{tabular}
\end{ruledtabular}\label{tab:t1}
\end{table}

\begin{table*}
\caption{Thermodynamic quantities and parameters at each temperature. Pressures at saturation $P_{\rm sat}$ are taken from \cite{psattot}.
Surface tensions $\gamma$ and bulk liquid densities $\rho\sub{l}$ are obtained using the fitting functions from \citep{baidakov},
see appendix A for details.}
\begin{ruledtabular}
\begin{tabular}{ l c c c c c c c c c} $kT/ \epsilon $ &    
 $P_{\rm sat}$ & 
 $\gamma$ &
 $\rho\sub{l}$  &  
 B$_{2}/\sigma^{3}$&
 $\eta$  & 
 $\xi$   & 
 $r\sub{c}$ 
\\ & $ [\epsilon/\sigma^3]$&$ [\epsilon/\sigma^2]$ &  $[m/\sigma^3]$& & & & $[\sigma]$ 
\\  \hline     
 1.0  &$2.55 \times 10^{-2}$    & 0.453 & 0.696 & 
   -5.26 & 2.79 & 1.94  & 1.26 \\ \hline
 0.8  &$4.53 \times 10^{-3}$   & 0.863 & 0.797 & 
   -7.75 & 6.07 & -1.52  & 1.33 \\\hline
 0.6    &$2.54 \times 10^{-4}$   & 1.33 & 0.882 & 
   -12.9 & 11.6 & -6.21 & 1.41 \\\hline
 0.5    &$2.54 \times 10^{-5}$   & 1.57 & 0.921 & 
   -18.15& 16.1&-9.46& 1.46 \\\hline
 0.4  & $8.02\times 10^{-7}$    & 1.83 & 0.959 
 & -28.8 & 22.8 & -13.9  & 1.52\\\hline
 0.3  & $2.53\times 10^{-9}$    &2.10 & 0.996 
  & -58.2 & 33.9 & -20.7  & 1.60\\
\end{tabular}
\end{ruledtabular}\label{tab:t2}
\end{table*}

\section{Results}\label{sec:results}

\subsection{Nucleation rates}

\begin{figure}[h]
  \includegraphics[width=.38\textheight]{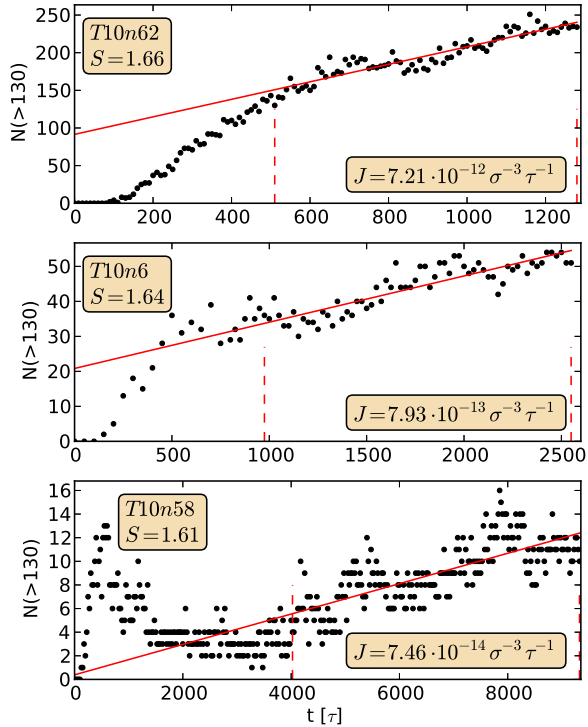}
  \caption{Number of clusters above the threshold size as a function of time. The nucleation rate is the slope of the linear fit (solid line). An initial lag time was ignored 
  for these fits, the end of the lag time is given by the first of the vertical dashed lines. See Table \ref{tab:t3} for the 1-$\sigma$ errors 
on the nucleation rate.}\label{fig:T10}
\end{figure}

\begin{figure}
  \includegraphics[width=.4\textheight]{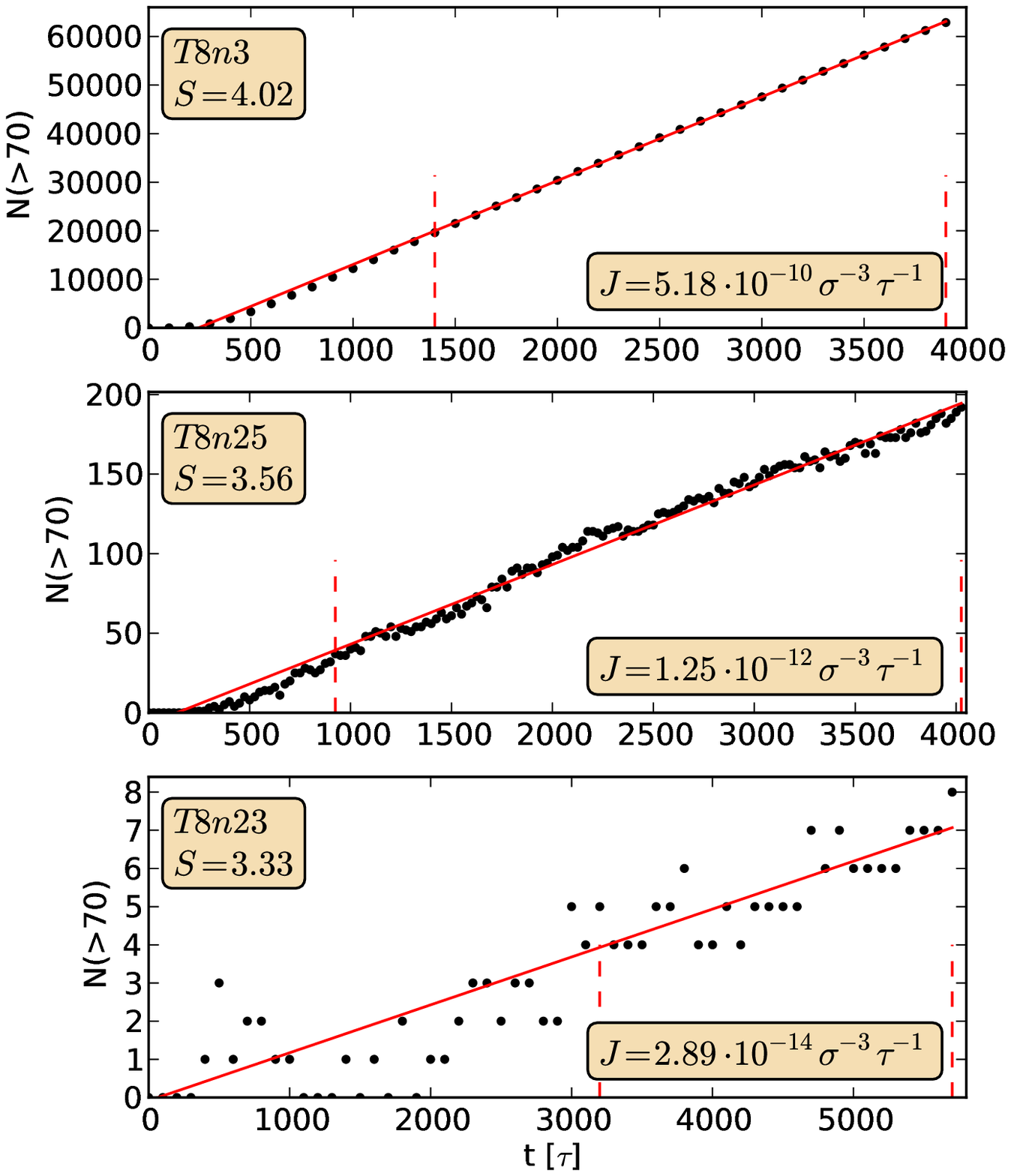}
  \caption{Like figure \ref{fig:T10}, but for T = 0.8 $\epsilon/k$ .}\label{fig:T8}
\end{figure}

\begin{figure}
  \includegraphics[width=.38\textheight]{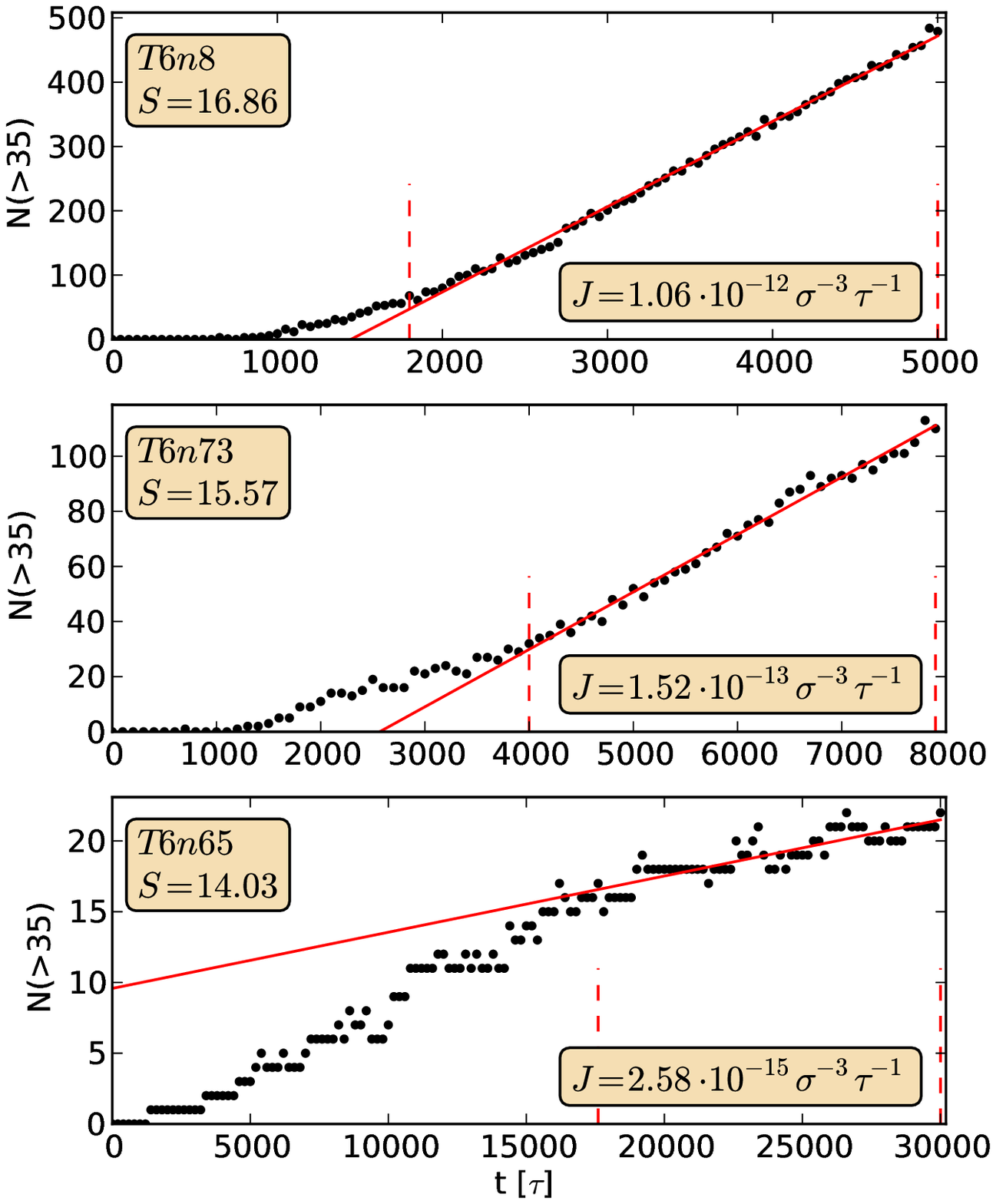}
  \caption{Like figure \ref{fig:T10}, but for T = 0.6 $\epsilon/k$ .}\label{fig:T6}
\end{figure}

\begin{figure}
  \includegraphics[width=.38\textheight]{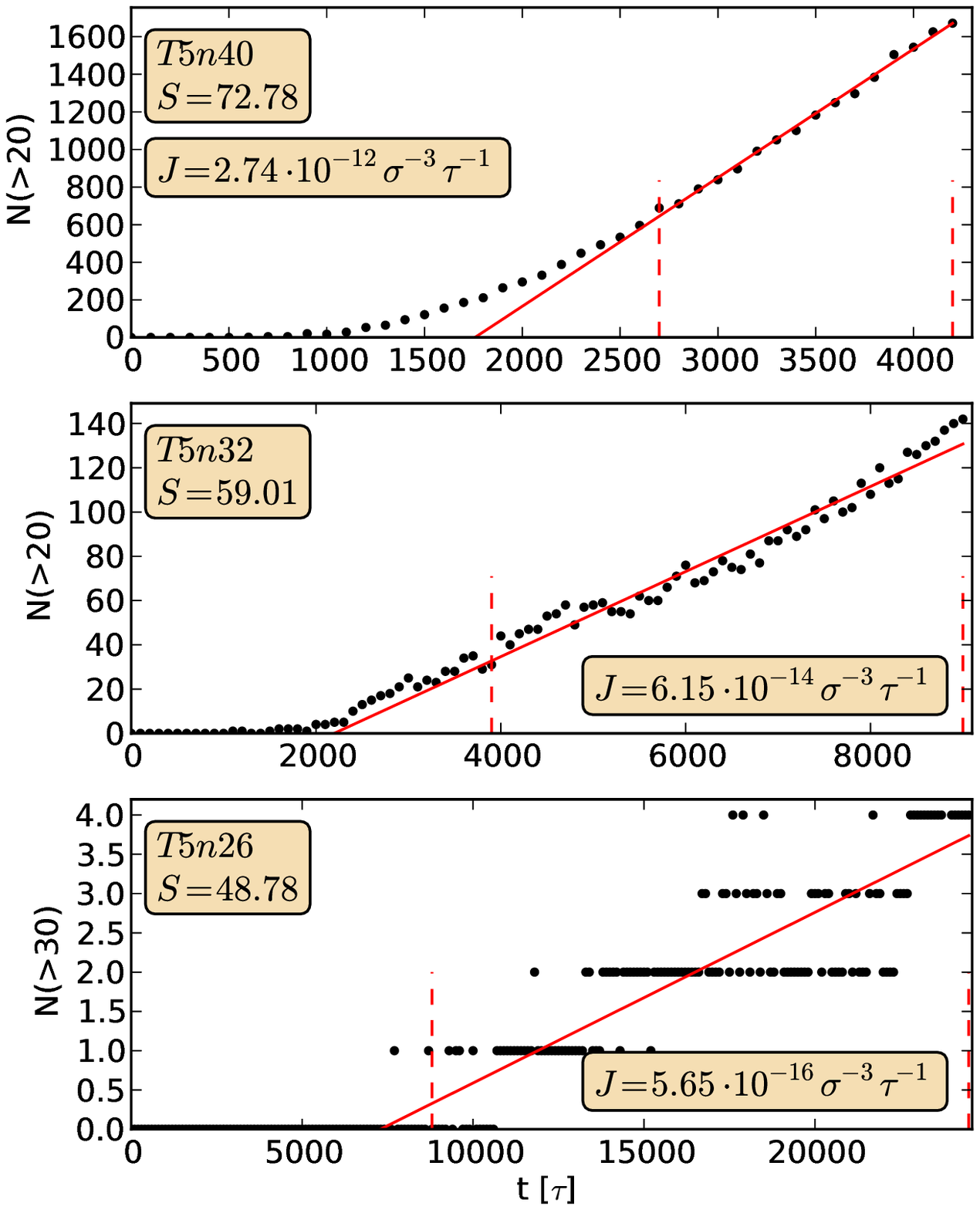}
  \caption{Like figure \ref{fig:T10}, but for T = 0.5 $\epsilon/k$ .}\label{fig:T5}
\end{figure}

\begin{figure}
  \includegraphics[width=.38\textheight]{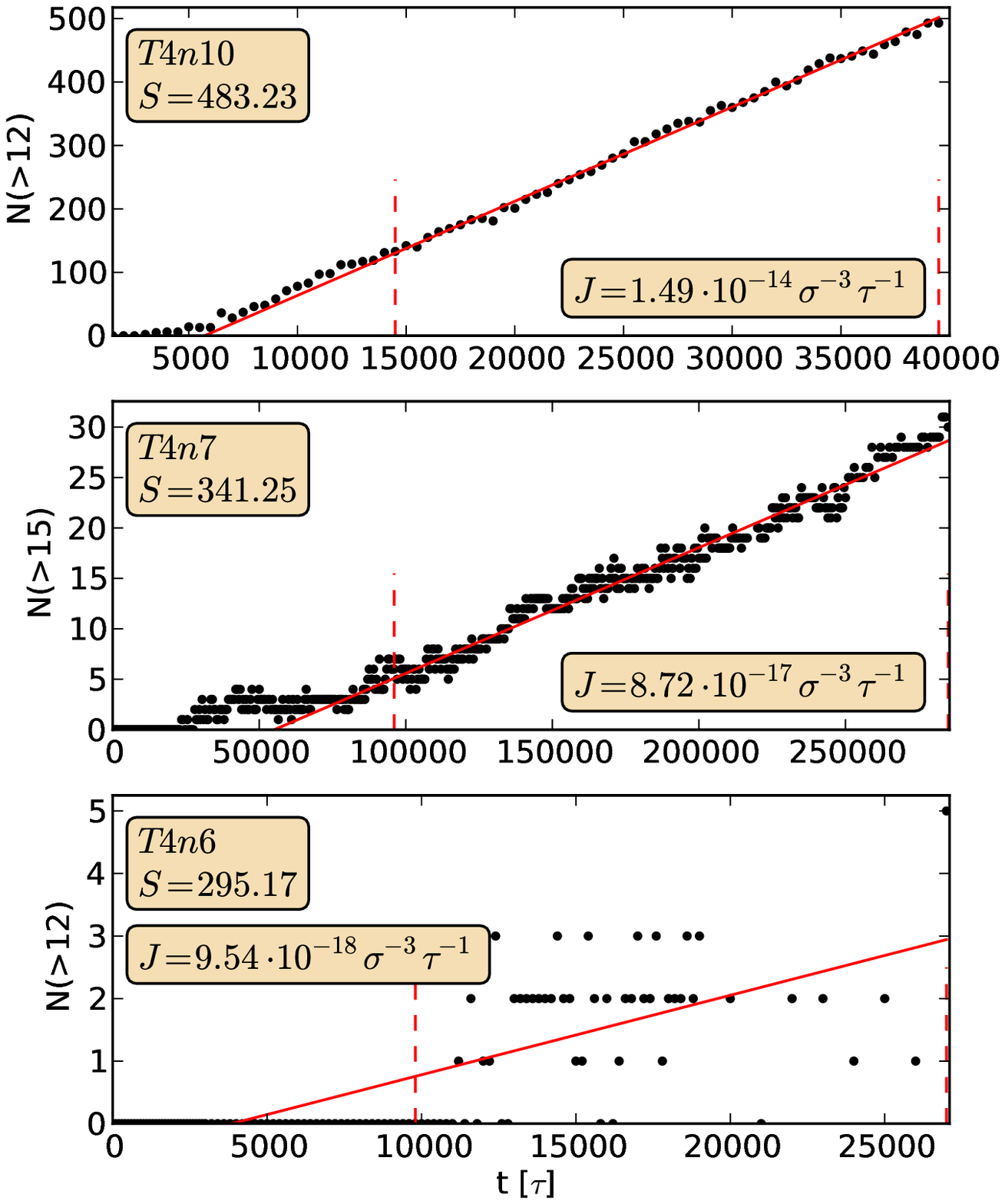}
  \caption{Like figure \ref{fig:T10}, but for T = 0.4 $\epsilon/k$ .}\label{fig:T4}
\end{figure}

\begin{figure}
  \includegraphics[width=.38\textheight]{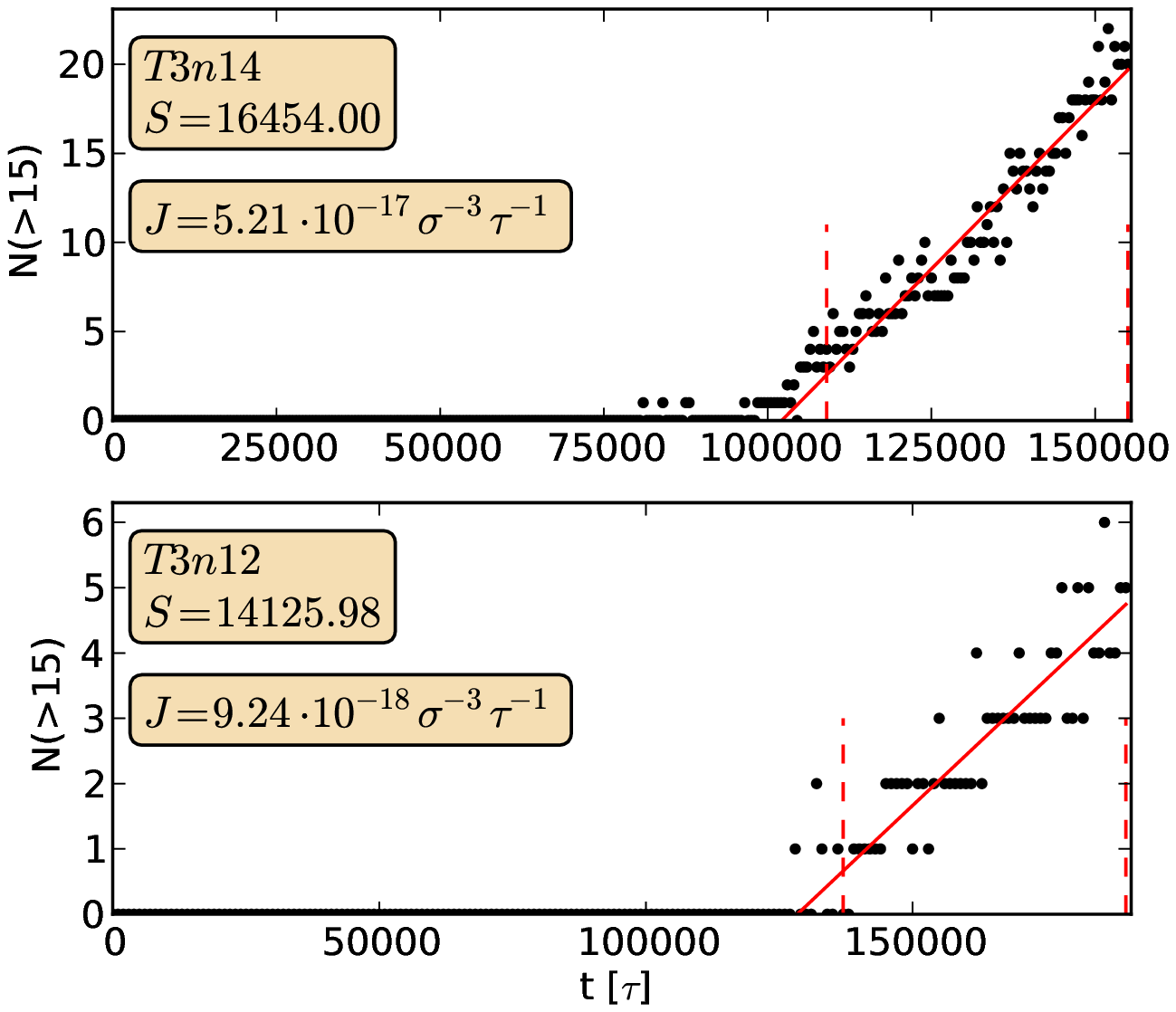}
  \caption{Like figure \ref{fig:T10}, but for T = 0.3 $\epsilon/k$ .}\label{fig:T3}
\end{figure}

\begin{figure}
\includegraphics[width=.38\textheight]{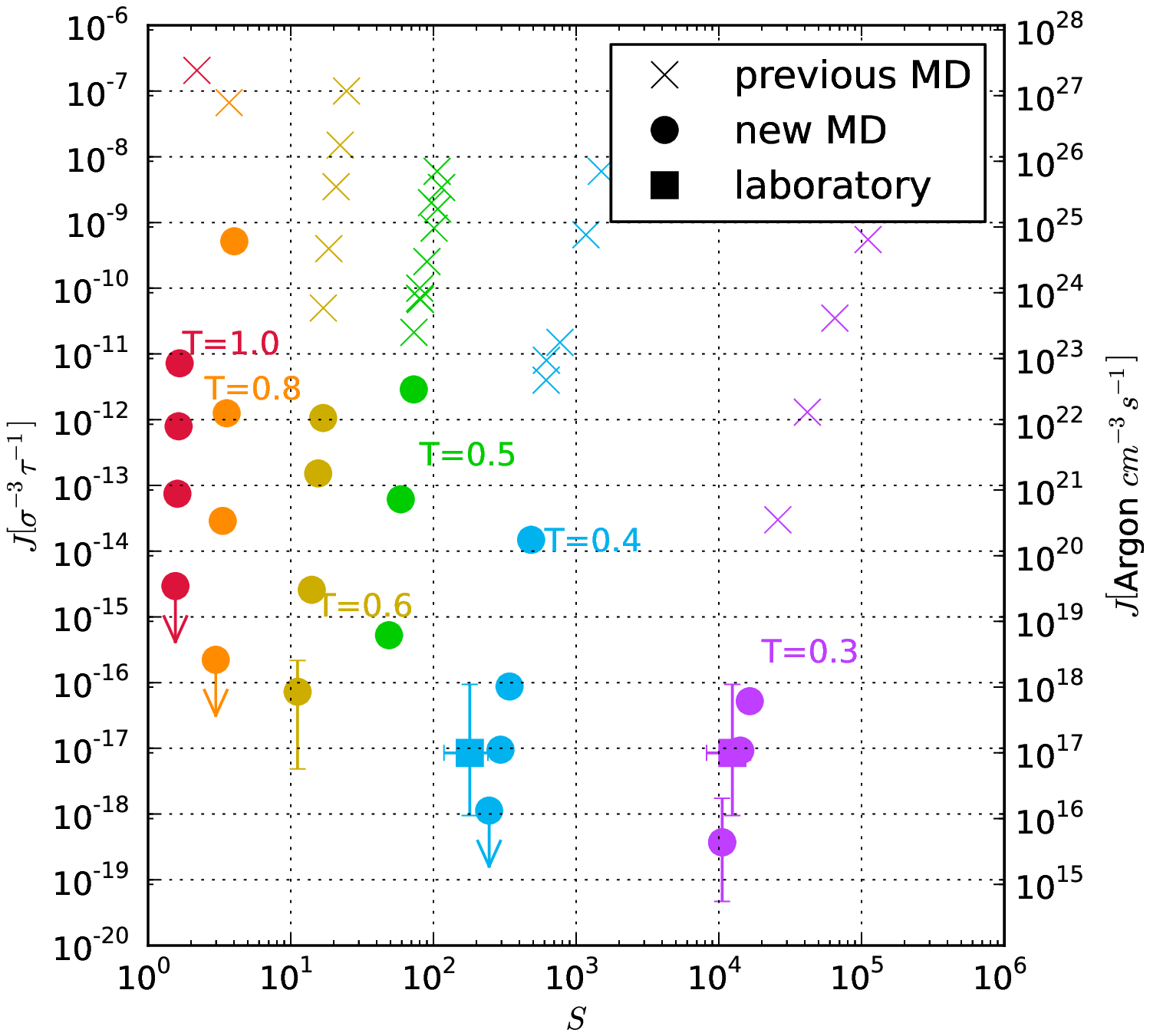}
\caption{Overview of the nucleation rates measured in the MD simulation 
presented here (circles and arrows) and from Tanaka \textit{et al.} 2005, Tanaka \textit{et al.} 2011 and Wedekind 2007 \cite{Tanaka2005,Tanaka2011, wedekind} (crosses). Uncertainties are smaller than
the symbols for most of our runs (see Table \ref{tab:t3}) and not are not displayed here, with the exception of runs T6n55 and T3n9,
where the observation of only one stable cluster leads to the large 68 percent confidence ranges indicated by the error bars.
Runs which did not nucleate were used to derive upper limits (68 percent confidence levels) on the nucleation rates, plotted with downward arrows.
The two squares mark recently-measured Argon nucleation rates from experiment \cite{sinha}.}\label{fig:Joverv}
\end{figure}

Nucleation rates $J$ are derived from the rate at which the number of clusters above some threshold size grows. For example, in T8n3 (figure \ref{fig:T8}), the increase in the amount of clusters possessing at least 70 members, is, after normalization by the simulation volume, the nucleation rate. This is always measured after some initial lag time, visible in the figure as the first vertical red dashed line. The rates are found to be independent of size threshold, as long as the threshold is larger than $i^*$, while the lag times increase with threshold size.
See for example, Figure \ref{fig:T8n3}, which shows the evolution of $N(>i)$ for a wide range of threshold sizes. Unlike smaller simulations, our runs do not run out of gas: Because there is no significant vapor depletion over the nucleation phase, our nucleation rates remain constant throughout the runtime. In smaller simulations, drops in the nucleation rates are seen as soon as a significant number of
stable clusters have formed. This due to the resulting drop in monomer pressure and supersaturation ratio (see e.g. Figure 3 in Tanaka \textit{et al.}\cite{Tanaka2011}, or Figure 2 in Chkonia \textit{et al.}\cite{chkonia}). In such cases, the time interval for measuring the nucleation rate must be carefully chosen. We are able to simply use the entire simulation time period after the initial lag time for measuring nucleation rates. 

Figures \ref{fig:T10} to \ref{fig:T3} show the data and least squares
linear fits used to derive the nucleation rates. The uncertainties in $J$ come from the standard deviation of the slope in the least square fit and also from
our choice for the lag time cutoff. The second was estimated by allowing the lag times
to vary by 10 percent around our chosen values (the ones given by vertical dashed lines in Figures \ref{fig:T10} to \ref{fig:T3}) and measuring width of the range containing 68 percent of the best fit slopes. These two errors were added in quadrature to give the total 68\% error margins reported in Table \ref{tab:t3}. 

The runs can be categorised roughly according to the abundance of stable clusters by the end of the run:
\begin{itemize}
\item \textbf{Numerous nucleation events} 
Most runs form a large number of stable clusters and allow us to measure $J$ accurately, with uncertainties as low as a few percent. 

\item \textbf{Moderate nucleation events}
Runs T3n12, T4n6 and T5n26 show nucleation, however because the rates are so low, few stable clusters are produced.  The number of time-steps required to reach the constant slopes for $N(>i)$ for a range of threshold sizes $i$ (as illustrated in Figure \ref{fig:T8n3}) becomes large. The slopes of $N(>i)$ in the available time period depend on the choice for the threshold size $i$, as well as the assumed initial lag time. This, as well as the low number of stable clusters leads to increased uncertainties in the nucleation rates.
\item \textbf{Few or no nucleation events}
T6n55 and T3n9 formed one stable cluster: The probability for this is 16\% for rates yielding 2.90 and 0.195 stable clusters on average, resulting the wide 68\% confidence interval in Table \ref{tab:t3}.

Runs T10n55, T8n2 and T4n5 have no stable clusters by the end of the simulations. Their nucleation rates lie beyond our available computational resources.
The upper limits on their nucleation rates in Table \ref{tab:t3} were derived from the Poisson distribution, which implies that with 68 percent confidence the nucleation rate lies between zero and a $J$ value
which on average produces 1.14 stable clusters per simulation volume during the nucleation period
(which was assumed to last for 90 percent of the run time, to account for some unknown initial lag time).
A higher confidence, 90 percent, upper limit lies at 2.02 times the upper limits given in Table \ref{tab:t3}.

T4n6 produces few clusters, making an accurate determination on the nucleation rate difficult. Comparison with the well-resolved higher supersaturation run T4n7 shows that the correct $J$ can be measured using a time
interval of $\tau = (1.0 - 2.7)\times 10^4$, and a threshold size of $i=11$ or larger. The models suggest that $i^*(T4n6) \simeq i^*(T4n7) + 1$. Therefore $i=12$ should give a good estimate
for $J$ for run T4n6, even though this run did not reach a stage of mature nucleation where the slopes of $N(>i)$ are exactly constant and independent of $i$ over a wide range in $i$.
\end{itemize}
Refer to table \ref{tab:t3} for a comprehensive list of results. 

\subsection{Critical sizes from the first nucleation theorem}

From Eqs. (\ref{eqdist}), (\ref{eq:j}) and (\ref{MCNT}) one can derive the first nucleation theorem\cite{Oxtoby,Kalikmanov2013}: 

\begin{equation}\label{nucltheo}
i^*_{\rm NT} = \left( \frac{\partial \ln J}{\partial \ln S} \right)_T   - 1 \; .
\end{equation}

It allows us to derive the critical cluster sizes $i^*$ from the nucleation rates $J_{\rm{MD}}$.
We estimate the derivative by taking the finite differences to the next available nucleation rate at the same temperature. If both a higher and a lower rate are available these two rates are used to calculate the slope. Some runs showed no nucleation events
and only give upper limits on the nucleation rates. Together with the next higher nucleation rate measurement at the same temperature, they set a lower limit on the derivative and on $i^*_{\rm{NT}}$.

At low temperatures, $kT \le 0.6\epsilon$, the critical sizes from the nucleation theorem agree quite well with Eq. (\ref{icrit}), i.e. with the peak position in $\Delta G_{\rm CNT}$ and $\Delta G_{\rm MCNT}$, see Table \ref{tab:t3}.
Good agreement between $i^*_{\rm NT}$ and $i^*_{\rm CNT}$ at low temperatures (45 - 70K) was also found in LJ MD simulations at higher $S$ and $J>10^{23}$ cm$^{-3}$s$^{-1}$ by Wedekind et. al\cite{Wedekind2007} and in LJ MC calculations at $T = 0.741\epsilon/k$ [\cite{wolde}].
The SP-model underestimates the critical sizes by a small amount at low temperatures (however its nucleation rate predictions are much more accurate than those from CNT and MCNT).
At $kT \ge 0.8\epsilon$ both CNT and the SP model underestimate the critical sizes significantly.

\subsection{Comparison with model predictions}\label{sec:nuclrates}

\begin{figure}[h]
\includegraphics[width=.38\textheight]{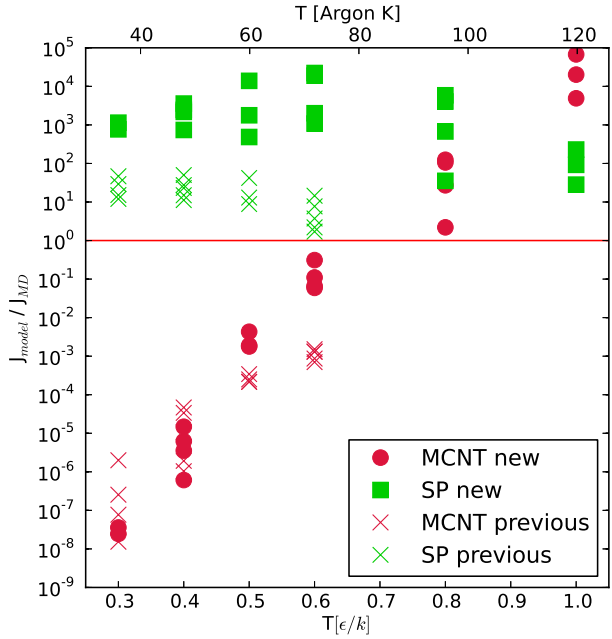}
\caption{Comparison of the nucleation rates from MD simulations with the predictions from the MCNT and SP models.
Model predictions are compared to the MD simulations presented here (square and circles) and to previous 
higher supersaturation MD simulations (crosses)\cite{Tanaka2005,Tanaka2011}.}
\label{fig:Jmodels}
\end{figure}

Nucleation rates predicted by the SP and MCNT models are given in Table \ref{tab:t3} and they are plotted in comparison with our MD nucleation rate measurements in Figure \ref{fig:Jmodels}. Nucleation
rates measured in earlier, higher supersaturation MD simulations \cite{Tanaka2005,Tanaka2011} are also compared to these two models in the same way.
The CNT model is not shown in this comparison - it predicts significantly smaller nucleation rates than the MCNT model
and is known to differ from simulations and experimental results by large factors \cite{Tanaka2005}.

\begin{table*}\caption{Total pressure $P$ measured in the simulation, supersaturation $S$ (pressure $P$  divided by
the saturation pressure $P_{\rm sat}$), critical cluster size $i^*$, nucleation rate $J$ and sticking probability $\alpha$ for each run.
The critical sizes $i_{\rm{NT}}^*$ were derived form the measured rates $J_{\rm{MD}}$ using the first nucleation theorem, Eq. (\ref{nucltheo}).
The nucleation rates were derived using the MCNT and SP model (with $\alpha = 1.0$) and measured directly in the MD simulations.
Also included are nucleation rate predictions from a hybrid model $J_{\rm hybrid}$ (using $\alpha_{\rm MD}$), see section \ref{sec:hybrid} for details.}
\begin{ruledtabular}
\begin{tabular}{l |cc|ccc|cccc |c  }
  Run ID & $P $&  $S $ &
$i_{\rm{NT}}^*$& $i_{\rm{SP}}^*$& $i_{\rm{CNT}}^*$  &
$J_{\rm{MD}}$ &
$J_{\rm{SP}}$ &
$J_{\rm{MCNT}}$& $J_{\rm{\rm hybrid}}$& $\alpha_{\rm MD}$\\
 & $[\epsilon/\sigma^3]$ & & & & & $\left[ \sigma^{-3}\tau^{-1}\right]$&$\left[ \sigma^{-3}\tau^{-1}\right]$&$\left[ \sigma^{-3}\tau^{-1}\right]$& $\left[ \sigma^{-3}\tau^{-1}\right]$&
\\ 
\hline 
T10n62&4.24$\times10^{-2}$&1.66&129&62&49&7.21$\pm$0.06$\times10^{-12}$&2.04$\times10^{-10}$&3.53$\times10^{-8\phantom{0}}$&1.70$\times10^{-11}$&0.077\\

T10n60&4.17$\times10^{-2}$&1.63&126&68&54&7.93$\pm$0.83$\times10^{-13}$&7.35$\times10^{-11}$&1.62$\times10^{-8\phantom{0}}$&2.24$\times10^{-12}$&0.061\\

T10n58&4.09$\times10^{-2}$&1.60&108&76&60&7.46$\pm$0.80$\times10^{-14}$&1.73$\times10^{-11}$&5.09$\times10^{-9\phantom{0}}$&2.49$\times10^{-13}$&0.046\\

T10n55&3.96$\times10^{-2}$&1.55&$>$99&93&75&$<$1.10$\times10^{-14}$&1.13$\times10^{-12}$&5.71$\times10^{-10}$&-&-\\ \hline

T8n30&1.82$\times10^{-2}$&4.02&48&21&25&5.27$\pm$0.02$\times10^{-10}$&1.487$\times10^{-8}$&9.21$\times10^{-10}$&6.33$\times10^{-10}$&0.19\\

T8n25&1.61$\times10^{-2}$&3.55&51&29&32&1.25$\pm$0.02$\times10^{-12}$&8.46$\times10^{-10}$&3.39$\times10^{-11}$&1.35$\times10^{-12}$&0.11\\

T8n23&1.51$\times10^{-2}$&3.33&49&34&38&3.38$\pm$0.26$\times10^{-14}$&1.17$\times10^{-11}$&2.90$\times10^{-12}$&1.04$\times10^{-13}$&0.10\\

T8n20&1.35$\times10^{-2}$&2.98&$>$45&46&51&$<$2.00$\times10^{-16}$&1.05$\times10^{-12}$&1.90$\times10^{-14}$&-&-\\ \hline

T6n80&4.29$\times10^{-3}$&16.9&24&15&21&1.09$\pm$0.01$\times10^{-12}$&7.02$\times10^{-10}$&3.80$\times10^{-14}$&9.73$\times10^{-13}$&0.16\\

T6n73&3.96$\times$10$^{-3}$&15.6&32&16&23&1.53$\pm$0.04$\times10^{-13}$&1.85$\times10^{-10}$&6.11$\times10^{-15}$&7.71$\times10^{-14}$&0.13\\

T6n65&3.57$\times10^{-3}$&14.0&38&18&25&2.58$\pm$0.19$\times10^{-15}$&2.97$\times10^{-11}$&4.84$\times10^{-16}$&2.18$\times10^{-15}$&0.12\\

T6n55&3.04$\times10^{-3}$&11.95&21$-$40&23&31&0.49$-$7.21$\times10^{-17}$&9.76$\times10^{-13}$&4.83$\times10^{-18}$&5.65$\times10^{-18}$&0.088\\ \hline

T5n40&1.85$\times10^{-3}$&72.8&18&10&16&2.74$\pm$0.14$\times10^{-12}$&7.10$\times10^{-10}$&2.60$\times10^{-15}$&4.56$\times10^{-12}$&0.24\\

T5n32&1.50$\times10^{-3}$&59.2&20&12&18&6.15$\pm$0.18$\times10^{-14}$&5.53$\times10^{-11}$&5.86$\times10^{-17}$&5.57$\times10^{-14}$&0.14\\

T5n26&1.24$\times10^{-3}$&48.7&23&14&21&5.26$\pm$0.3$\times10^{-16}$&7.31$\times10^{-12}$&2.26$\times10^{-18}$&5.70$\times10^{-16}$&0.10\\ \hline

T4n10&3.88$\times10^{-4}$&484&12&9&15&1.49$\pm$0.01$\times10^{-14}$&4.31$\times10^{-12}$&8.34$\times10^{-20}$&1.44$\times10^{-14}$&0.21\\

T4n7&2.74$\times10^{-4}$&342&14&12&18&8.99$\pm$0.3$\times10^{-17}$&7.13$\times10^{-14}$&1.55$\times10^{-22}$&2.73$\times10^{-17}$&0.13\\

T4n6&2.37$\times10^{-4}$&295&15&12&19&9.54$\pm$2.42$\times10^{-18}$&1.30$\times10^{-14}$&1.21$\times10^{-23}$&7.37$\times10^{-19}$&0.06\\

T4n5&1.97$\times10^{-4}$&246&$>$10&14&21&$<$1.01$\times10^{-18}$&8.17$\times10^{-16}$&2.06$\times10^{-25}$&-&-\\ \hline

T3n14&4.17$\times10^{-5}$&16460&13&8&13&1.32$\pm$0.05$\times10^{-16}$&1.22$\times10^{-14}$&5.53$\times10^{-25}$&7.08$\times10^{-17}$&0.16\\

T3n12&3.58$\times10^{-5}$&14130&14$\pm$2&8&13&1.56$\pm$0.08$\times10^{-17}$&3.23$\times10^{-15}$&7.00$\times10^{-26}$&7.66$\times10^{-18}$&0.13\\

T3n9&2.69$\times10^{-5}$&10620&15$\pm$5&9&15&5.3$-$100$\times10^{-20}$&1.97$\times10^{-16}$&8.98$\times10^{-28}$&8.14$\times10^{-20}$&0.09\\
\end{tabular}
\end{ruledtabular}\label{tab:t3}

\end{table*}

\begin{itemize}
\item \textbf{MCNT} At low temperatures ($T \le 0.6\epsilon/k$) the MCNT predictions lie below the measured rates $J_{\rm MD}$ by many orders of magnitude. At $T = 0.3\epsilon/k$ and $T = 0.4\epsilon/k$ the discrepancy is larger at the lower supersaturations simulated in this work, while at $T = 0.6\epsilon/k$ the trend goes in the opposite direction. The too-low nucleation rates predicted by the MCNT model are due
to its too-high $\Delta G(i)$ curves, which lead it to underestimate the equilibrium abundance of clusters near $i^*$ by large factors. Refer to Figure \ref{fig:dist-total}. 
$J_{\rm{MCNT}} / J_{\rm{MD}}$ increases strongly with temperature and at $T= 1.0 \epsilon/k$ the MCNT rates lie about
five orders of magnitude {\it above} the simulation values - the temperature dependence of $J$ in the MCNT model differs greatly from the simulation results. 
 
\item \textbf{SP} The SP model on the other hand matches the results from previous, smaller MD simulations\cite{Tanaka2005,Tanaka2011} at higher $S$ and $J$ quite well.
However, at the lower supersaturations probed here, the SP model overestimates the true rates significantly at all temperatures:
$J_{\rm SP}/J_{\rm MD}$ exceeds $10^4$ in some cases. This limitation of the SP model also results in deviations in the predictions for the equilibrium abundances
of small clusters, see Figure \ref{fig:dist-total}.

\end{itemize}

\subsection{Comparison with the Argon SSN experiment}\label{sec:experiment}

Most laboratory measurements of Argon nucleation probe nucleation
rates lower than
$10^9$ cm$^{-3}$ s$^{-1}$ (e.g. Iland \textit{et al.}\cite{Iland2007}). The recent
development of Laval Supersonic Nozzle (SSN)
nucleation experiments\cite{Sinha2008} has increased the accessible
rates enormously, by almost 10 orders of
magnitude. Together with the decrease in accessible $J$ rates by over
10$^4$ reached in the MD simulations
presented here, direct comparisons of experiments and MD simulations
are now possible.

\begin{figure}[!h]
\includegraphics[width=.38\textheight]{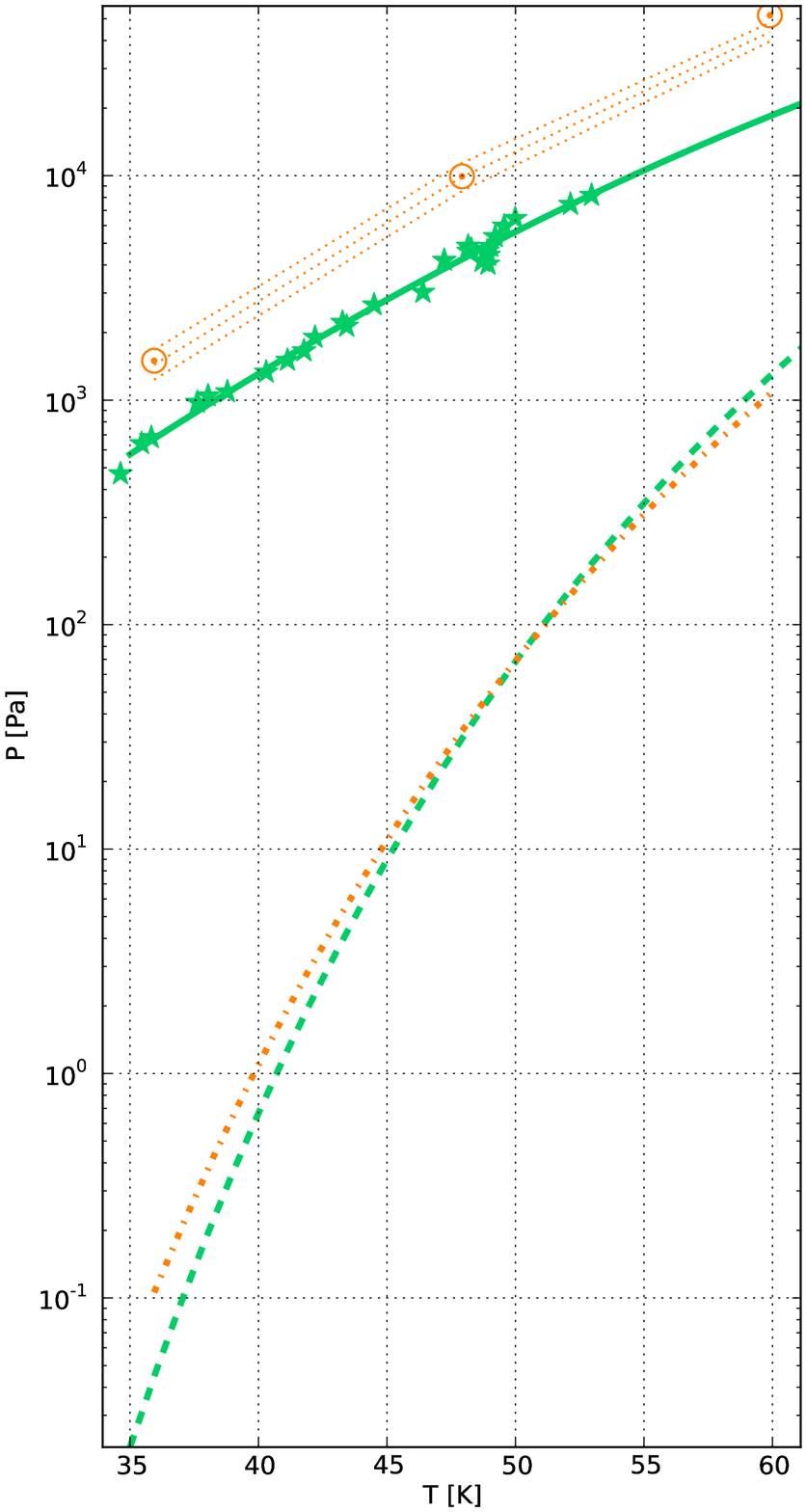}
\caption{Pressures and temperatures corresponding to nucleation rates of $J=10^{17\pm1}$ cm$^{-3}$
from the argon SSN experiment\cite{sinha} (green stars) and from our LJ MD simulations (dotted circles). The dashed green line
shows the extrapolated saturation pressures (i.e. the vapor/liquid equilibrium curve) assumed for argon\cite{sinha}, while the orange dash-dotted line shows
the curved used in this study for the LJ fluid.}
\label{fig:pvst_ssn}
\end{figure}

For the case of argon, SSN experiments have been performed in the
temperature range
of 34 to 53K at nucleation rates of $10^{17\pm1}$ cm$^{-3}$s$^{-1}$ [\cite{sinha}].
The temperatures and nucleation rates coincide directly with two of our simulations:
Run T3n12 at $T=36$K and
run T4n6 at $T=48$K both have nucleation rates close to $10^{17}$ cm$^{-3}$s$^{-1}$,
assuming the widely used argon system, where $\sigma=3.405$\AA, and $\epsilon/k = 119.8$K  $[$\cite{Michels1949,kalikmanovReview,Tanaka2011}].
Figure \ref{fig:pvst_ssn} directly compares simulations and experiments with in the pressure -- temperature plane: The LJ-fluid nucleates
at the same rate at pressures about 2.3 times above those
found in the argon experiment, indicating that the two substances have quite similar volatilites
and that a simple LJ model describes the nucleation properties of low temperature argon quite well.

For a more detailed comparison we convert the pressures to supersaturations. This requires extrapolation of the saturation pressures of argon and the LJ-fluid 
far below their well constrained temperature range. Both saturation curves have uncertainties of about 50 percent at these low temperatures. A LJ saturation curve a factor of
2.3 above the Argon curve is not ruled out at these low temperatures, and would lead to perfect agreement between experiment and simulation. 
To convert the pressures measured in the SSN experiment, we adopt the same argon saturation curve as in\citep{sinha}.
The resulting supersaturations closely follow the scaling relation
from Hale\cite{Hale1986}, see equations (7) and Fig. 6 in Sinha \textit{et al.}\citep{sinha}.
With Eq. (7) from \cite{sinha} one finds $S(T=0.3\epsilon/k)
= 12'430$ and $S(T=0.4\epsilon/k) = 180$, which we plot in Figure \ref{fig:Joverv} with estimated uncertainties in $S$ of 50 percent. 
At $T=0.3 \epsilon/k$ simulations and experiment agree very well,
while at $T=0.4 \epsilon/k$ the simulations require about 1.6 times
larger supersaturations to reach similar nucleation rates.
More accurate low temperature saturation curves for argon and for the LJ fluid are required to
determine if this difference is real, within the current large uncertainties experiment and simulations agree quite well.

Combining SSN with NPC data at $J = 10^{7\pm2}$cm$^{-3}$ s$^{-1}$ and using the first nucleation theorem, Eq. (\ref{nucltheo}) allows us to estimate critical sizes in a temperature range of 42 to 52 K. At 48 K the
result is $i^* \simeq 17\pm6$ [\cite{sinha}]. This agrees very well with our values of $i^*_{\rm NT} = 15$ and $i^*_{\rm CNT} = 19$ for run T4n6, which has the same nucleation rate as the SSN experiment.

The temperature scaling of nucleation rates relative to MCNT seems to be qualitatively different:
the simulations show an increasing discrepancy with the classical
nucleation rate predictions as the temperature is lowered (see Figure \ref{fig:Jmodels}
and also earlier simulation results\cite{Wedekind2007,Tanaka2011} and the NPC argon experiment\cite{Iland2007}).
In the SSN experiment this discrepancy is nearly constant or even slightly decreasing towards lower
temperatures (see Figure 8 in Sinha \textit{et al.}\cite{Sinha2008}). However, the pressure scaling is very similar
and the different supersaturation scaling is caused by
the different, and quite uncertain, slopes of the argon and LJ saturation curves used here (Figure \ref{fig:pvst_ssn}) .

Also note that the critical temperature for a LJ fluid is $T_c = 1.313(1)
\epsilon/k$, both for the potential\cite{Potoff1998,Perez-Pellitero2006,McGrath2010}
and for a cutoff at $5\sigma$ [\cite{ShiJohnson2001}].
The Argon critical temperature $T_c=150.80$\cite{webbook} implies a lower
conversion factor of $\epsilon/k = 114.85 K$. This would shift the
experimental data in Figure \ref{fig:Joverv} to the right, to
$S(T=0.3\epsilon/k) = 28'740$ and to$S(T=0.4\epsilon/k) = 300$.
Now the agreement at $T=0.4\epsilon/k$ would be perfect, and at the lowest temperature the experimental rate would be on the low side,
but still within the 50 percent error bars in S for both argon and the LJ fluid. 

\subsection{Cluster size distributions}

Below the critical size $i^*$ the abundance of clusters is stationary. It can be predicted by assuming a certain model for $\Delta G(i)$ and using 
the equilibrium distribution given by Eq. (\ref{eqdist}). The simulations give a cluster size distribution at every analyzed snapshot. To reduce
the statistical noise we take the time-averaged size distribution. To exclude the initial lag time, which is required to reach the stationary 
size distribution, we conservatively include only the second half of the simulated time period in each run for the time-averaged size distribution.
In model estimates, the number density $n(i)$  in the steady state is expressed in terms of the equilibrium number distribution $n\sub{e}(i)$, as in Tanaka \textit{et al.}\cite{Tanaka2011} 
\begin{eqnarray}
n(i)= J n_{\rm e}(i) 
\sum_{j=i}^{\infty}
{ 1 \over R^{+}(j) n_{\rm e}(j)}.
\label{ni-stationary}
\end{eqnarray} 

\begin{figure}[]
\includegraphics[width=.35\textheight]{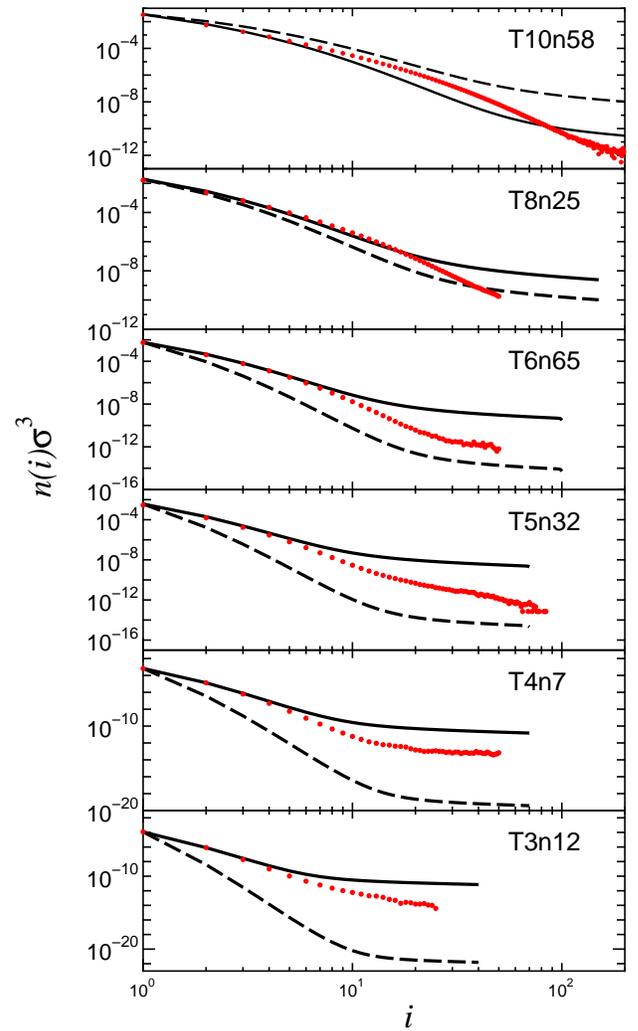}
\caption{Size distributions measured in some of our MD simulations (circles) compared with the predictions from the MCNT (dashed lines) and SP (solid lines) models.
The simulated size distributions are time averages over the second half of the simulated period.}
\label{fig:dist-total}
\end{figure}

Figure \ref{fig:dist-total} shows that the MCNT model underestimates the cluster abundance at low temperatures ($T \le 0.6\epsilon/k$) by large factors. This is directly related and agrees with the too-low nucleation rate predictions from the MCNT model in this temperature range.
The SP model matches very well for small cluster sizes, especially at low temperatures. This is why it manages to provide rather accurate nucleation rate estimates at high supersaturations, when
the critical cluster sizes are small.
At larger cluster sizes however, the SP model often over-predicts the equilibrium abundances. This explains the too-high nucleation rate estimates the SP model produces at low supersaturations, where the critical cluster are larger.

The comparison of observed and predicted size distributions (Figure \ref{fig:dist-total}) well illustrates the limited range in cluster size and temperature
where the theoretical models roughly match simulations. These limited ranges of validity are consistent with the discrepancies in predicted and observed
nucleation rates. (See section \ref{sec:nuclrates}).

\subsection{Free energy for cluster formation}\label{sec:free-energy-sec}

\begin{figure}
\includegraphics[width=.35\textheight]{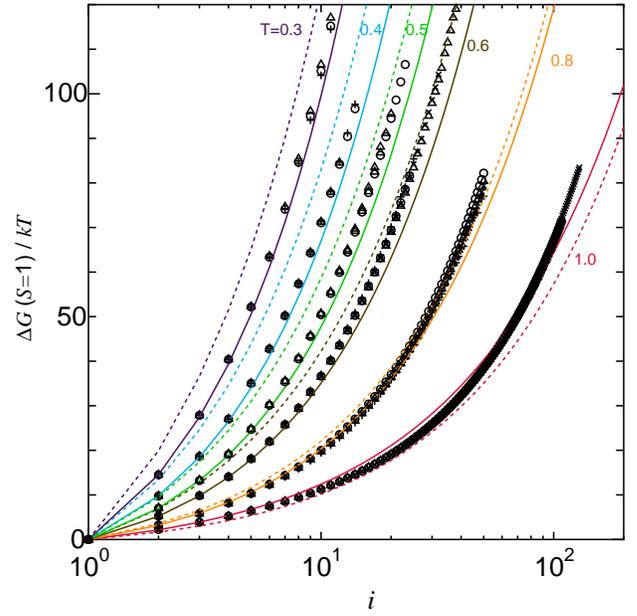}
\caption{The free  
energies at $S=1$  associated with forming a cluster for various 
temperatures. Different
symbols indicate $\Delta G_{i}(S=1)$ 
obtained from the MD simulations starting from different
supersaturation ratios. The predictions by the SP model (solid lines)
and by the modified-CNT (dotted lines) are also shown.
}\label{fig:deltag-total}
\end{figure}

The number density of clusters $n(i)$ in the
steady state is almost equal to $n_{\rm e}(i)$, for $i \siml
i^*$. Using the cluster size distributions from simulation, we
can infer the free energy of subcritical clusters:
\begin{equation}\label{free-energy}
\Delta G_{\rm MD} = -kT  \ln \left\{ n(i) \over P_1/(kT) \right\}.  
\end{equation} 
This, combined with the free energy supersaturation dependence
given by equations (\ref{MCNT}) and (\ref{SP}) gives us the
free energy at equilibrium ($S=1$):  
\begin{equation}\label{free-energy_s1}\nonumber
\Delta G_{\rm MD}(S=1) = -kT  \ln \left\{ n(i) \over P_1/(kT) \right\}+
(i-1) kT \ln S.  
\end{equation} 
Fig.~\ref{fig:deltag-total} shows $\Delta G_{\rm MD}(S=1)$ for all
runs. Only small clusters with $i<i^*_{\rm NT}$ are plotted, where
$i^*_{\rm NT}$ is the critical cluster size given by the first nucleation theorem.
This figure confirms that $\Delta G_{\rm MD}(S=1)$ depends only on 
temperature.  Predictions for $\Delta G (S=1)$ from the models
are also plotted here. The MCNT fails to correctly predict $\Delta G_{\rm
MD}(S=1)$ over all temperatures.  The SP model however fares better in matching the free
energy $\Delta G_{\rm MD}(S=1)$ for small clusters, especially at
low temperatures ($T \le 0.6 \epsilon/k$). For larger clusters the free energy curve of the SP model lies below with MD results,
which explains the too large nucleation rates it predicts in the low $J$ regime simulated in this work.

\subsection{Droplet growth and sticking probabilities}\label{sec:sticking}

\begin{figure}
\includegraphics[width=.38\textheight]{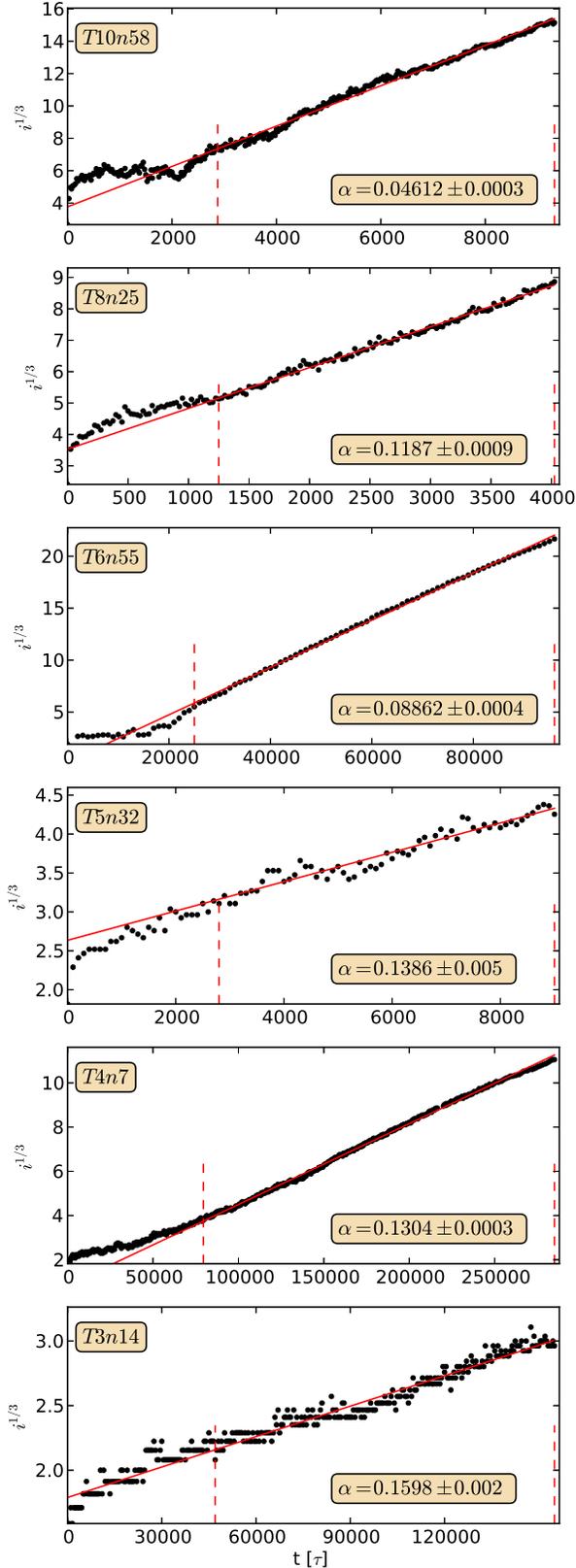}
\caption{The sizes of the largest clusters in the simulations are cubic with time. From these we can estimate the net growth rate per encounter $\alpha$, see Eq. (\ref{alpha}).} \label{fig:alpha_for_5_runs}
\end{figure}

\begin{figure}
\includegraphics[width=.38\textheight]{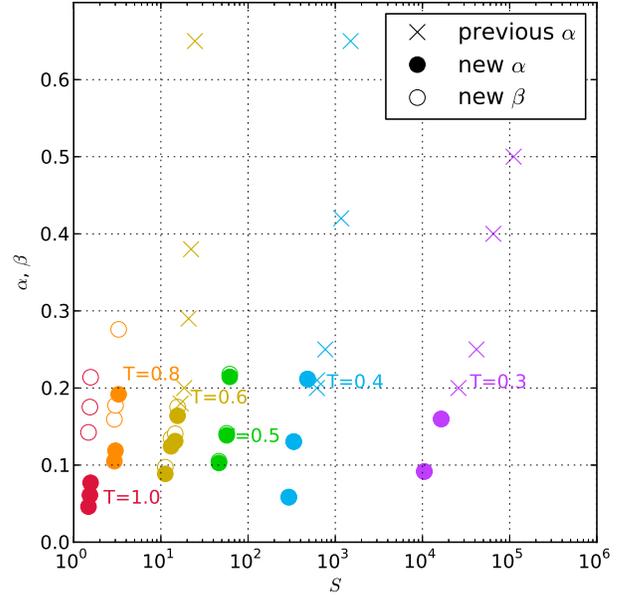}
\caption{The growth rates $\alpha$ in our simulations (solid circles) continue on the trend of decreasing $\alpha$ for decreasing supersaturation, as found previously\cite{Tanaka2011} (crosses).
For low supersaturations (high temperatures) evaporation is expected to be significant and the sicking probabilities $\beta$ (open circles) lie well above the net growth rates.} \label{fig:sticking_probability_overview}
\end{figure}

We can measure the growth rates $di/dt$ in the MD simulations and have used them to define $\alpha$, the growth rate per encounter in Eq. (\ref{alpha}).
Subtracting an evaporation estimate, see Eqs. (\ref{beta}) and (\ref{kk12}), also allows us to estimate the sticking probability $\beta$, the probability that a cluster - monomer encounter results in the accretion of the monomer:
\begin{equation}\label{betamd}
\beta \simeq \frac{3}{4\pi r_0^2\nu_{\rm th} n_1}\frac{d \left(i^{1/3}\right)}{dt}  \left[ 1 - \frac{1}{S}\right] ^{-1} \; .
\end{equation}
In each run, we use the largest cluster to estimate $i^{1/3}$. As expected\cite{Tanaka2011}, after the initial lag phase, we find $i\left(t\right)$ for clusters to be strongly cubic ($\propto t^3$) in all simulations, as plotted
for a few simulations in Figure \ref{fig:alpha_for_5_runs}. This indicates that the $\alpha$ and $\beta$ do not depend on the cluster size.
At a fixed temperature, the $\alpha$ values continue on the trend of decreasing $\alpha$ for decreasing $S$,
as found previously\cite{Tanaka2011}. Figure \ref{fig:sticking_probability_overview} gives an overview of the growth rates pre encounter and sticking probabilities for all simulations which formed stable clusters. 

The model predictions in Figure \ref{fig:Jmodels} assume $\alpha=1$, as usual. Using the measured $\alpha$ values instead, would lower the $J_{\rm model}$ values, but not nearly enough to make the SP model match the measured $J_{\rm MD}$ values.

\section{Hybrid nucleation model} \label{sec:hybrid}

Here we propose a new empirical nucleation model, which combines results from theory and simulation.
As shown in equations (1) and (\ref{eq:j}), 
the nucleation rate is determined by 
the equilibrium number density of clusters, $n_{\rm e}(i)$.
Using the free energy obtained in Section \ref{sec:free-energy-sec}, 
we can evaluate $n_{\rm e}(i)$. 
We set
\begin{eqnarray}\nonumber
\Delta G = 
\left\{ \begin{array}{ll} 
-(i-1)\ln S
+ \Delta G_{\rm MD}(i,S=1)   
 &,\; i \le i_{\rm  T}\\
 -(i-1)\ln S +
  \eta (i^{2/3}-1)kT + D
 &, \; i > i_{\rm  T},   \\
\end{array} \right.
\label{empirical-deltag}
\end{eqnarray} 
where $D$ is defined so that $\Delta G$ is continuos at the transition scale $i_{\rm  T}$
\begin{equation}
D=  \Delta G_{\rm MD}(i_{\rm  T},S=1)
 - \eta ( i_{\rm  T}^{2/3}-1)kT \;.
\end{equation} 
In other words, the constant D is the difference between the free energy functions of the hybrid model and the MCNT model
above the transition scale $i_{\rm  T}$.

For the evaluation of $\Delta G_{\rm MD} $, 
we take the size distributions from low $J$ runs T10n58, T8n25, T6n65, T5n26, T4n7, and T3n9. At each temperature, we set $i_{\rm  T}$ to the critical
size from the first nucleation theorem, $i_{\rm{NT}}^*$, evaluated for these six runs. Table \ref{tab:t3} lists nucleation-rate estimates for this model for all our runs, using the
growth rate per encounter $\alpha$ as measured from simulation.

\begin{figure}
\includegraphics[width=.35\textheight]{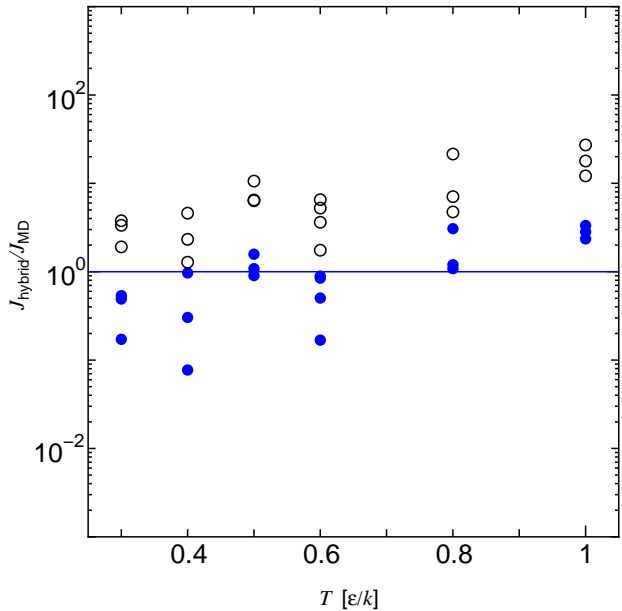}
\caption{The ratios between the nucleation rates $J_{\rm hybrid}/J_{\rm MD}$.
For the points indicated by the open circles, the sticking probability $\alpha$ is assumed to be unity, whereas the values of $\alpha$ obtained by MD simulations are designated by the filled circles.}\label{fig:jratio-hybrid}
\end{figure}

The ratios of the nucleation rates between the hybrid model
and the MD simulations are plotted in Fig. \ref{fig:jratio-hybrid} for two cases: one in which  $\alpha= 1$ and the
other in which $\alpha$ is set to the value obtained directly from 
simulation. By taking into account the realistic $\alpha$-values, we find that 
the hybrid model agrees with the simulations within one order of magnitude
for all cases. 

The relative success of this hybrid approach in comparison to purely theoretic strategies helps pinpoint shortfalls in the standard model pictures. In the standard model framework, the free energy of subcritical clusters can be obtained from the subcritical cluster distribution - this, under the assumption of equilibrium - via Eq. (\ref{eqdist}). Because nucleation is a non-equilibrium process, this Boltzmann distribution might not be accurate for clusters close to $i^*$.  That the hybrid model succeeds to match the simulated rates quite well implies that the free energies from Eq. (\ref{eqdist}) are quite accurate even for clusters almost as large as $i^*$. The hybrid model relies on the volume term form classical models, $(i-1)\ln S$, its success indicates that this term is indeed correct .(However we had to define the $S$ using the total pressures to get meaningful results at
$T = 1.0 \epsilon/k$, while in the theoretical models the monomer pressures are used.)
The failings of the purely theoretical models therefore are contained within the surface term contributions to $\Delta G$.

\section{Summary}\label{sec:summary}

We have performed large scale MD simulations of homogeneous vapor to liquid nucleation for a wide range of temperatures and supersaturations, using up to eight billion Lennard-Jones atoms. In this paper we present the first results form these simulations, which are summarized as follows:
\begin{itemize}
\item The large scale of the MD simulations presented here allows us to accurately measure nucleation rates as low as $10^{-17}\left[ \sigma^{-3}\tau^{-1}\right]$
 and form critical cluster sizes $> 100$. The supersaturation in such large volume simulations remains practically constant during nucleation, the rates are independent of time and threshold size
 and can be measured very precisely. A large number of stable and unstable droplets are formed under realistic conditions, their microscopic properties will be presented in a
 subsequent paper (Ang\'elil et al. in preparation). 
\item The simulated nucleation rates allow for a direct comparison with the SSN argon experiment\cite{sinha}: Within the uncertainties, we find good agreement in the pressures and
supersaturations required to nucleate at a rate of $10^{17}$ cm$^{-3}$ s$^{-1}$: Our pressures are about 2.3 times higher. At $36$K the supersaturations agree perfectly,
 while at $48$K it is about 1.6 lower in the experiment. Further studies of the LJ fluid and argon at these low temperatures are required in order to quantify the significance of these small differences.
\item We confirm that classical models (CNT and MCNT) severely underestimate nucleation rates at low temperature, and that the discrepancy becomes larger for lower temperatures.
At $T = 0.8 \epsilon/k$ the rates are quite accurate and at $T = 1.0 \epsilon/k$ they are overestimated by up to $10^5$.
The critical size predictions match the values from the nucleation theorem quite well at low temperatures, and are too low at $T\ge 0.6 \epsilon/k$.
\item The SP model, which matches MD simulation nucleation rates at higher supersaturations quite well\cite{Tanaka2005,Wedekind2007,Tanaka2011}, overestimates
 the rates in the regime probed here significantly at all temperatures. Its critical size predictions are generally too low, especially at high temperatures.
\item The growth rate of clusters above the critical size is exactly proportional to $i^{2/3}$, which confirms that growth rates per encounter do not depend on the droplet size\cite{Tanaka2011}. The growth rates per encounter continue to decrease towards lower supersaturations. We measure values as low as $\alpha = 0.046$.
  Accurate nucleation rate estimates at low supersaturations depend on $\alpha$-values. For lower supersaturations than probed here, they could be obtained from
  MD simulations by following the growth of existing, super-critical liquid clusters embedded in low supersaturation vapor.
\item We present a hybrid nucleation model, which mixes elements from CNT with simulation results: In the free energy function it uses an empirical surface term derived from the subcritical
 cluster abundance in the simulations, combined with a classical volume term. The predicted rates from the hybrid model agree well with the simulations. This suggests that the classical
 framework for modelling nucleation could work quite well, if some non-classical, more accurate surface term is used in the free energy function.
\end{itemize}

\section{Acknowledgments}
We thank the referee Vitaly Shneidman and a second referee for detailed and constructive reports. We acknowledge a PRACE award (36 million CPU hours) on Hermit at HLRS.
Additional computations were preformed on SuperMUC at LRZ, on Rosa at CSCS and on the zBox4 at UZH.
J.D. and R.A. are supported by the Swiss National Science Foundation. The authors thank B. Hale, J. Hutter, V. Kalikmanov, A. Laio, D. Reguera and I. Siepmann for discussion and comments.  
\appendix

\section{Model parameters}\label{sec:model_details}

As a contribution to the free energy $\Delta G(i),$ all three nucleation models (Eqs. \ref{CNT}, \ref{MCNT} and \ref{SP}) include a surface term $\propto \eta$. $\eta$ is related to $\gamma,$
the planar surface tension of the condensed phase via Eq. (\ref{eq:eta}).
For the surface tension $\gamma$, we use the fitting function \cite{baidakov}
\begin{equation}
\gamma = 2.942 \times \left(1 - T / T_c \right)^{1.303} \; , T_c = 1.312 \epsilon/k \; , 
\end{equation}
which matches the available simulation results\cite{psattot,Dunikov2001,baidakov,Barrett2008} well. 
$r_0$ depends on the bulk density $\rho_m,$ and is defined by
\begin{equation}
r_0 = \left(\frac{3m}{4\pi\rho_m}\right)^\frac{1}{3}.
\end{equation}
In the same study the bulk density of the LJ liquid is parametrized by \citep{baidakov},
\begin{equation}\nonumber
\rho_m = 0.0238\cdot \left(13.29 + 24.492 f^{0.35} + 8.155 f\right) \;\; \left[m/\sigma^3\right],
\end{equation}
with 
\begin{equation}
f = 1 - \frac{T}{1.257 \left[\epsilon/k\right]}.
\end{equation}
In addition to a surface term which depends on $\eta,$ the SP model incorporates one which depends on $\xi$. This parameter can be set with
\begin{equation}\nonumber
\xi = -\frac{1}{2^{1/3}-1}\left[\ln\left(\frac{-B_2 P_{\rm sat, 1}}{kT} \right) +\left(2^{2/3} - 1\right)\eta \right],
\end{equation}
where $P_{\rm sat,1}$ is the saturation pressure of the monomer gas component, which we estimate from the total saturation pressure
using the virial expansion. $B_2$ the second virial coefficient given by
\begin{equation}\label{B2}
B_2 = 2\pi \int^\infty_0 \left(1-\exp{\left[-\frac{u\left(r\right)}{kT}\right]}\right)r^2dr,
\end{equation} 
with $u\left(r\right)$ the Lennard-Jones potential (Eq. \ref{lj}), which we cut off and shift to zero at $r=6.78\sigma$. To be consistent with the other thermodynamic
quantities we use same cutoff scale in Eq. (\ref{B2}) as used in\citep{baidakov}, instead of the $5\sigma$ cutoff used in our simulations.
We find very similar nucleation rates with this longer cutoff as with the $5\sigma$ cutoff, see Section \ref{sec:conv}.

Based on MD simulation results\cite{psattot}, the saturation pressure of a Lennard-Jones liquid can be parametrized by
\begin{equation}\label{psattot}
P_{\rm sat} = \frac{\epsilon}{\sigma^3}\exp{ \left[A - \frac{C\epsilon}{kT}\right]},
\end{equation}
with coefficients $A = 3.24157$, and $C = 6.91117$. This relation fits MD results\cite{psattot,baidakov} in the range $0.5 \le kT/\epsilon \le 1.2$
and also Monte Carlo calculations\cite{Barrett2008} in the range $0.25 \le kT/\epsilon \le 0.875$. Note that the uncertainties in $P_{\rm sat}$
are about a factor of 2 at our lowest temperature $kT/\epsilon = 0.3$.

\section{The effect of evaporation}
\label{sec:evaporation}

The evaporation rate is obtained from the principle of detailed balance in the thermal equilibrium:
\begin{equation}\label{balance}
R^-(i+1)n_e(i+1) = R^+(i)n_e(i)  \; .
\end{equation}
Combined with Eqs. (\ref{growth}) and (\ref{rplus}), we have
\begin{equation}
\frac{di}{dt} = \beta n_e(1) \nu_{\rm th} \; 4\pi r_0^2\left[ i^{2/3}  - \frac{n_e(i-1)}{n_e(i)} (i-1)^{2/3}  \right],
\end{equation}
and using equilibrium number densities $n_e$ from Eq. (\ref{eqdist}) one finds
\begin{equation}\label{kk7}
\frac{di}{dt} = \beta n_e(1) \nu_{\rm th} \; 4\pi r_0^2\left[ i^{2/3}  - \frac{e^{\Delta G(i)} }{e^{\Delta G(i-1)} } (i-1)^{2/3} \right]  .
\end{equation}

To evaluate the evaporation term one has to assume a certain form for $\Delta G$. Here we use the CNT form from Eq. (\ref{CNT})
for simplicity (the other models considered in this work, MCNT and SP, lead to the same conclusion)
\begin{equation}
\frac{e^{\Delta G_{\rm CNT}(i)} }{e^{\Delta G_{\rm CNT}(i-1)}} = \frac{ e^{\eta i^{2/3} - \eta(i-1)^{2/3} } }{S}.
\end{equation}
For large clusters $(i \gg 1)$, $ (i-1)^{2/3}  \simeq i^{2/3}$ and Eq. (\ref{kk7}) reduces to this simple approximation:
\begin{equation}\label{kk12}
\frac{di}{dt} = \beta n_e(1) \nu_{\rm th} \; 4\pi r_0^2 i^{2/3} \left[ 1  - \frac{1}{S} \right]  .
\end{equation}
This suggests that for large clusters $R^-(i) \simeq R^+(i)/ S$ and that evaporation plays a significant role at low supersaturations.


%

\end{document}